\documentclass[11pt]{article}
\usepackage{a4wide}
\usepackage{amsmath,amssymb}
\usepackage{simplewick}
\usepackage{graphicx}
\usepackage{color}
\usepackage{braket}
\usepackage{comment}
\usepackage[bookmarks=true,bookmarksnumbered=true,setpagesize=false]{hyperref}

\usepackage{fancybox}
\usepackage{url}

\DeclareMathOperator{\sgn}{sgn}
\DeclareMathOperator{\erf}{erf}

\DeclareMathOperator{\Res}{Res}

\usepackage{enumerate}

\usepackage{fontawesome}
\newcommand{\fs}{\text{\faStarO}}

\newcommand{\E}{\textbf{\textsf E}}
\newcommand{\M}{\textbf{\textsf M}}
\newcommand{\T}{\textbf{\textsf T}}
\newcommand{\DelT}{\Delta\!\T}

\newcommand{\sr}{\stackrel}

\newcommand{\al}[1]{\begin{align}#1\end{align}}

\newcommand{\ov}{\over}
\newcommand{\nn}{\nonumber\\}
\newcommand{\tx}{\text}

\newcommand{\paren}[1]{\left(#1\right)}
\newcommand{\pn}{\paren}
\newcommand{\sqbr}[1]{\left[#1\right]}
\newcommand{\ab}[1]{\left|#1\right|}
\newcommand{\br}[1]{\left\{#1\right\}}
\newcommand{\fn}[1]{\!\left(#1\right)}

\newcommand{\Paren}[1]{\bigl(#1\bigr)}
\newcommand{\Pn}{\Paren}
\newcommand{\Sqbr}[1]{\bigl[#1\bigr]}

\newcommand{\Fn}[1]{\!\bigl(#1\bigr)}

\newcommand{\bs}{\boldsymbol}

\newcommand{\df}{\text{d}}

\newcommand{\green}[1]{{\color[cmyk]{0.97,0,0.75,0}#1}}

\newcommand{\mc}{\mathcal}
\newcommand{\mf}{\mathfrak}

\newcommand{\p}{\partial}

\newcommand{\Or}[1]{\mathcal O\!\left(#1\right)}

\newcommand{\ol}{\overline}

\newcommand{\wh}{\widehat}

\newcommand{\vs}{\varsigma}

\newcommand{\stin}{\varsigma_\tx{in}}
\newcommand{\stout}{\varsigma_\tx{out}}

\newcommand{\s}{\textsf s}

\begin{document}

\title{
New effect in wave-packet scattering of quantum fields 
\bigskip}

\author{
Kenzo~Ishikawa,$^*$
Kenji~Nishiwaki,$^\dagger$
and
Kin-ya~Oda$^\ddag$\bigskip\\
$^*$ \it\normalsize Department of Physics, 
Hokkaido University, Hokkaido 060-0810, Japan\smallskip\\
$^*$\it\normalsize Research and Education Center for Natural Sciences, Keio University Kanagawa 223-8521, Japan\smallskip\\
$^\dagger$ \it\normalsize Department of Physics, 
Shiv Nadar Institution of Eminence, Gautam Buddha
Nagar 201314, India\smallskip\\
$^\ddag$ \it\normalsize Department of Physics, 
Osaka University, Osaka 560-0043, Japan\smallskip\\
$^\ddag$ \it\normalsize Department of Mathematics, Tokyo Woman’s Christian University, Tokyo 167-8585, Japan\bigskip
}

\maketitle


\begin{abstract}\noindent
We report calculations of a wave-packet amplitude of the two-body scattering $\phi \phi \to \Phi \to \phi \phi$, which leads to the measured probability in realistic experiments. We elucidate the decay amplitude of $ \Phi \rightarrow \phi \phi$ from this.
In such an amplitude of wave packets, there are in and out time boundaries for the initial $\Phi$ and final $\phi\phi$ configurations, respectively.
In this paper, we prove that the effect of the in time boundary of $\Phi\to\phi\phi$ emerges from $\phi\phi\to\Phi\to\phi\phi$ without assuming any time boundary \emph{a priori}. This effect has been overlooked in the standard plane-wave formulation and can exhibit distinct phenomena in wide areas of science.
We confirm the result in different integration orders.
The result is also interpreted as a Stokes phenomenon in the Lefschetz-thimble decomposition.
\end{abstract}

\newpage
\tableofcontents

\newpage
\section{Introduction}
Particle scattering in quantum field theory requires wave packets in its very foundation, whereas the plane-wave formulation, which involves the square of the energy-momentum delta function in S-matrix~\cite{Schwartz}, is ``more a mnemonic than a derivation''~\cite{Weinberg:1995mt}.
More concretely, this means the following:
\begin{itemize}
\item Particle scattering in quantum field theory in standard calculational method uses plane waves. The amplitude of plane waves is proportional to a four-dimensional Dirac delta function ${\delta^{4}\fn{p_\tx{f}-p_\tx{i}}}$, showing energy-momentum conservation, where $p_\tx{i}$ and $p_\tx{f}$ are the initial- and final-state total momenta.  A square of the Dirac delta function is proportional to ${\delta^4\fn{0}}$ and is divergent literally. A standard approximation is to replace the infinity with a time interval and spatial volume of the system and to compute the proportional constant. Scattering cross section and other physical quantities are computed in this way. These are idealistic quantities that preserve space-time and other symmetries.
\item Physical states in experiments and natural phenomena have finite sizes in reality. In cases where these sizes are much larger than typical characteristic lengths of scatterings, these may be approximated with infinity, and plane-wave amplitudes describe the transition with a good approximation.  In other cases, these may not be well approximated with plane wave amplitudes.  In both cases, the total probability of a particle (within the one-particle subspace) is unity and is expressed by a normalized wave, satisfying $\ab{\Braket{\psi|\psi}}^2=1$. These are the wave packets.  Scattering amplitudes of normalized wave packets describe these transition processes and may become different from those of plane waves. 
\end{itemize}

Historically computations of scattering amplitudes have been developed by Tomonaga~\cite{Tomonaga:1946zz}, Schwinger~\cite{Schwinger:1949zz}, Feynman~\cite{Feynman:1949hz,Feynman:1949zx}, and Dyson~\cite{Dyson:1949bp} in the early time of quantum electrodynamics (QED); see Ref.~\cite{Schwinger:1958wrk} for related seminal papers.  Complicated calculations of higher-order corrections became drastically simplified by the powerful method of Feynman diagrams.  Amplitudes in the momentum space are expressed using two-point Green's functions,  propagators,  and vertex parts systematically.
In the context of this paper, it is important that Feynman dropped contributions from the asymptotic time region in order to obtain the beautiful Lorentz-invariant formula.
Feynman himself mentions the ignorance~\cite{Feynman:1949zx} when he proceeds {\it ``imagining that we can neglect the effect of interactions''} near the asymptotic time region, hoping that {\it ``we do not lose much in a general theoretical sense by this approximation''}; see Sec.~\ref{Wave-packet Feynman propagator section} for further details.
In the present paper, we calculate the amplitude including such a contribution neglected by Feynman.






The wave function in the asymptotic time region can be critical in a scattering process $\phi \phi \rightarrow \Phi \rightarrow \phi \phi$ of a light scalar $\phi$ via a heavy intermediate scalar $\Phi$ having a finite lifetime.
The wave-packet amplitude of this process is sensitive to the state in the asymptotic time region of $\Phi$.
This is difficult to analyze in the plane-wave formalism and is deeply connected with the contribution dropped by Feynman. The wave-packet amplitude takes into account this region explicitly and has no ambiguity nor difficulty in the computation.

In this paper, we present detailed calculations and the structure of the wave-packet amplitude.
We will show that such a contribution for $\Phi$ does exist and shows unusual and different properties not existing in the ordinary plane-wave Feynman diagram calculations.

Finally, we point out the possible phenomenological impact of our theoretical findings.
Ishikawa et al.\ claim that indeed a wave-packet effect---more specifically the \emph{time-boundary effect} due to localization of wave-packet overlap in time---is responsible for diverse phenomena in science such as the LSND neutrino anomaly~\cite{Ishikawa:2011qz,Ishikawa:2014eoa}; violation of selection rules~\cite{Ishikawa:2014uma}; the solar coronal heating problem~\cite{Ishikawa:2015gfa,ISHIKAWA2023100174}; anomalous Thomson scattering and a speculative alternative to dark matter, as well as modified Haag theorem~\cite{Ishikawa:2016lnn}; the anomalous excitation energy transfer in photosynthesis~\cite{Maeda:2017lbz}; and anomalies in the width of $e^+e^-\to\gamma\gamma$, in the $\pi^0$ lifetime, in Raman scatterings, and in the water vapor continuum absorption~\cite{Ishikawa:2019nes,Ishikawa2021}. 
There is an ongoing experimental project for this effect~\cite{Ishikawa:2019nes,Ushioda:2019hje}.

However, the time-boundary effect has not been paid high attention to, 
because so far it depends on whether one accepts \textit{a priori} the concept of the finite-time scattering that involves the time boundaries.
%
Here, we fill the gap by showing that the effect from the \emph{in time boundary} of the $\Phi\to\phi\phi$ decay~\cite{Ishikawa:2013kba,Ishikawa:2014eoa,Ishikawa:2018koj} \emph{emerges} from the $\phi\phi\to\Phi\to\phi\phi$ scattering amplitude~\cite{Ishikawa:2020hph} even if we do not include the in and out time boundaries for $\phi\phi$.
This way, we exhibit the necessity to include the time boundaries in general.


The organization of this paper is as follows:
In Sec.~\ref{Gaussian S-matrix section}, we review the basics of the Gaussian wave-packet formalism and show how to obtain the wave-packet S-matrix, namely the finite transition amplitude between the normalizable multi-wave-packet states.
In Sec.~\ref{emergence of time boundary section}, we show the emergence of the above-mentioned in time boundary for $\Phi$. 
In Sec.~\ref{Order 2 section}, we confirm our result by examining the pole structure of what we call the \emph{wave-packet Feynman propagator} and by referring to the Lefschetz thimble decomposition.
In Sec.~\ref{summary section}, we summarize our result.
In Appendix~\ref{notation section}, we list our notations.
In Appendix~\ref{simplifying section}, we present simpler expressions of S-matrix etc.\ for a particular configuration.
In Appendix~\ref{Details on Fig 2}, we give a detailed discussion on Fig.~\ref{illustration}.

\section{S-matrix in Gaussian formalism}\label{Gaussian S-matrix section}
We study the scattering $\phi\phi\to\Phi\to\phi\phi$ with an interaction Lagrangian density
\al{
\mc L_\tx{int}
	=	-{\kappa\ov2}\phi^2\Phi,
		\label{interaction Lagrangian}
}
where $\phi$ and $\Phi$ are real scalar fields with masses $m$ and $M$ ($>2m$), respectively, and $\kappa$ is a coupling constant.
We only take into account the tree-level $s$-channel scattering as we are mostly interested in the amplitude near the resonance pole of $\Phi$.

\begin{figure}\centering
\includegraphics[width=0.4\textwidth]{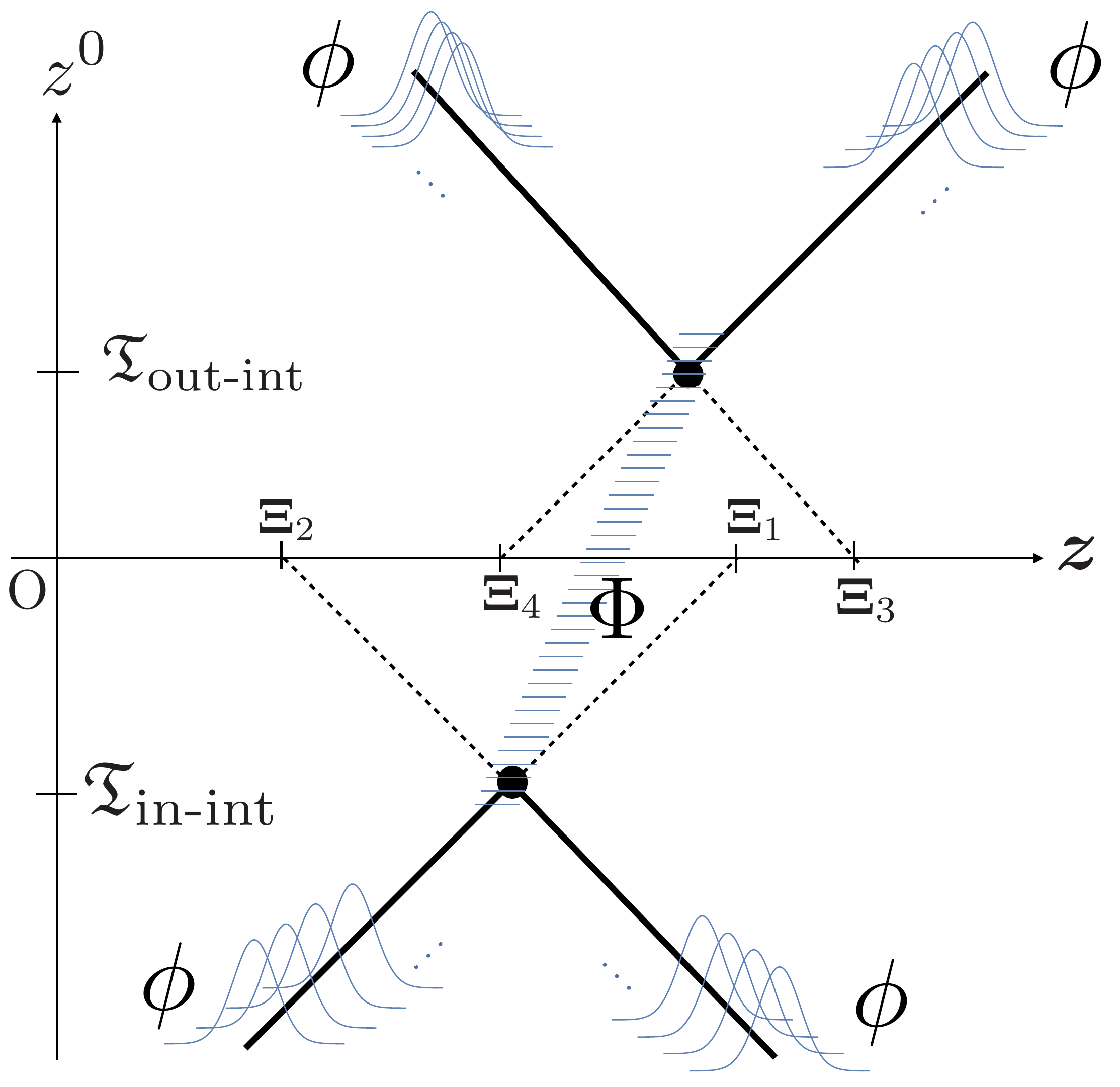}
\caption{Schematic figure in the position space $z=\pn{z^0,\bs z}$ for the intersection times $\mf T_\tx{in-int}$, $\mf T_\tx{out-int}$ and for the center of wave packet $\bs\Xi_a$ ($a=1,\dots,4$) at the arbitrarily chosen reference time $t=0$. The solid and dashed lines denote the trajectories of the wave packets of $\phi$. The intermediate state of $\Phi$ is a plane wave that spreads infinitely. The integral over the in- and out-interaction points $x=\pn{x^0,\bs x}$ and $y=\pn{y^0,\bs y}$ have largest support around the black dots at $z^0=\mf T_\tx{in-int}$ and $\mf T_\tx{out-int}$, respectively.
\label{schematic figure}}
\end{figure}

\subsection{Gaussian wave-packet formalism}
In this paper, we work in the Gaussian wave-packet formalism~\cite{Ishikawa:2005zc}, where the external fields $\phi$ are written in terms of the Gaussian basis, whereas the internal field $\Phi$ the plane-wave one;\footnote{
The result does not change if we use the Gaussian basis instead of the plane-wave one in internal lines; see footnote~\ref{comment on internal line}.
}
see e.g.\ Refs.~\cite{Ishikawa:2018koj,Ishikawa:2020hph,Oda:2021tiv} for reviews.
In Fig.~\ref{schematic figure}, we show a schematic figure.
The concrete form of the wave-function in the spacetime coordinate $z=\pn{z^0,\bs z}$ for the Gaussian wave packet of $\phi$ is, for particle labels $a=1,2,3,4$,
\al{
f_{\sigma_a;X_a,\bs P_a}\fn{z}
	&=	\left.
		\pn{\sigma_a\ov\pi}^{3/4}\int{\df^3\bs p\ov\sqrt{2p^0}\pn{2\pi}^{3/2}}
		e^{ip\cdot\pn{z-X_a}-{\sigma_a\ov2}\pn{\bs p-\bs P_a}^2}
		\right|_{p^0=\sqrt{m^2+\bs p^2}},
		\label{full Gaussian}
}where
%
the parameters for each wave packet $a$ (with $a=1,2$ for incoming and $3,4$ for outgoing) are its width-squared $\sigma_a$, spacetime position of its center $X_a=\pn{X_a^0,\bs X_a}$,
and its central momentum $P_a=\pn{E_a,\bs P_a}$, where $E_a:=(m^2+\bs P_a^2)^{1/2}$. We may trade $\bs P_a$ for $\bs V_a:=\bs P_a/E_a$ as independent parameters.

The tree-level $s$-channel S-matrix for the Gaussian wave packets~\eqref{full Gaussian} with the interaction~\eqref{interaction Lagrangian} becomes~\cite{Ishikawa:2020hph}
\al{
\mc S
	&=	\pn{-i\kappa}^2\int{\df^4p\ov\pn{2\pi}^4}{-i\ov p^2+M^2-i\epsilon}\nn
	&\quad\times
		\int\df^4x\,f_{\sigma_1;X_1,\bs P_1}\fn{x}f_{\sigma_2;X_2,\bs P_2}\fn{x}e^{-ip\cdot x}
		\int\df^4y\,f^*_{\sigma_3;X_3,\bs P_3}\fn{y}f^*_{\sigma_4;X_4,\bs P_4}\fn{y}e^{ip\cdot y}.
		\label{original S in Supplemental Material}
}
Here, the exponential factors originate from the plane-wave expansion of $\Phi$; the spacetime position of ``in-interaction'' $x$ and that of ``out-interaction''~$y$ are integrated over the whole spacetime.\footnote{\label{comment on internal line}
The result of the integral would not differ if we expanded $\Phi$ by the Gaussian waves instead of the plane waves~\cite{Ishikawa:2020hph}.
}

The amplitude~\eqref{original S in Supplemental Material} specifies the location of the wave packets $\bs X_1,\dots,\bs X_4$ in addition to their momenta $\bs P_1,\dots,\bs P_4$. One can integrate out $\bs X_1,\dots,\bs X_4$ if one wants to compare it with the corresponding plane-wave result. The amplitude~\eqref{original S in Supplemental Material} contains more information than the plane-wave S-matrix which only contains information of $\bs P_1,\dots,\bs P_4$.

We may rewrite the plane-wave propagator of $\Phi$ with the off-shell momentum $p=\pn{p^0,\bs p}$ as
\al{
{-i\ov p^2+M^2-i\epsilon}
={-i\ov-\pn{p^0}^2+\bs p^2+M^2-i\epsilon}
={-i\ov-\pn{p^0}^2+E_{\bs p}^2-i\epsilon}
={-i\ov-\pn{p^0}^2+\E_{\bs p}^2},
}
where $E_{\bs p}$ ($>\epsilon>0$) and $\E_{\bs p}$ are defined for later use:
\al{
E_{\bs p}
	&:=	\sqrt{M^2+\bs p^2},&
\E_{\bs p}
	&:=	\sqrt{E_{\bs p}^2-i\epsilon}\,
	\simeq
		E_{\bs p}-{i\epsilon\ov2E_{\bs p}}.
	\label{E_p defined}
}
Loop corrections to the two-point function of~$\Phi$ may be approximated by taking
\al{
\epsilon=M\Gamma,
}
where $\Gamma$ is the decay width of $\Phi$.\footnote{
See e.g.\ Appendix C and D in Ref.~\cite{Nishiwaki:2011gm} and references therein for subtleties when $\Gamma$ becomes comparable to $M$.
}
That is, within this particular theory, $\epsilon$ is not an infinitesimal nor an independent variable as it can be computed for a given set of parameters $\pn{\kappa,m,M}$. In this paper, we keep it independent for ease of extension.
To summarize, $\pn{\kappa, m, M, \epsilon}$ and $\pn{\sigma_a, X_a, \bs P_a}$ are all the independent parameters in this paper: The former set determines the theory,
while the latter set parametrizes each wave packet.

We also define the following for later use:
\begin{itemize}
\item $\sigma_\tx{in}$ and $\sigma_\tx{out}$ are the spatial widths-squared for the in and out interaction regions, respectively:
\al{
\sigma_\tx{in}
	&:=	{\sigma_1\sigma_2\ov\sigma_1+\sigma_2},&
\sigma_\tx{out}
	&:=	{\sigma_3\sigma_4\ov\sigma_3+\sigma_4}.
}
\item $\stin$ and $\stout$ are the temporal widths-squared for the in and out interaction regions, respectively:
\al{
\stin
	&:=	{\sigma_1+\sigma_2\ov\pn{\bs V_1-\bs V_2}^2},&
\stout
	&:=	{\sigma_3+\sigma_4\ov\pn{\bs V_3-\bs V_4}^2},
}
\item Their sum and the inverse of their inverse sum are\footnote{
In particular, ${1\ov\sigma}
	=	{1\ov\sigma_\tx{in}}+{1\ov\sigma_\tx{out}}=\sum_{a=1}^4{1\ov\sigma_a}$.
}
\al{
\sigma_+
	&:=	\sigma_\tx{in}+\sigma_\tx{out},&
\sigma
	&:=	{\sigma_\tx{in}\sigma_\tx{out}\ov\sigma_\tx{in}+\sigma_\tx{out}},
		\label{sigma plus}\\
\vs_+
	&:=	\vs_\tx{in}+\vs_\tx{out},&
\vs
	&:=	{\vs_\tx{in}\vs_\tx{out}\ov\vs_\tx{in}+\vs_\tx{out}}.
		\label{varsigma defined}
}
\item $\bs P_\tx{in}:=\bs P_1+\bs P_2$ and $\bs P_\tx{out}:=\bs P_3+\bs P_4$ are the total momenta for the in and out states, respectively. 
\item $E_\tx{in}:=E_1+E_2$ and $E_\tx{out}:=E_3+E_4$ are the total energies for the in and out states, respectively.
\item The following are the weight-averaged quantities:
\al{
\ol{\bs V}_\tx{in}
	&:=	\sigma_\tx{in}\pn{{\bs V_1\ov\sigma_1}+{\bs V_2\ov\sigma_2}},&
\ol{\bs V}_\tx{out}
	&:=	\sigma_\tx{out}\pn{{\bs V_3\ov\sigma_3}+{\bs V_4\ov\sigma_4}},&
\bs P_\sigma
	&:=	{\sigma_\tx{in}\bs P_\tx{in}+\sigma_\tx{out}\bs P_\tx{out}\ov\sigma_\tx{in}+\sigma_\tx{out}}.
	\label{P_sigma defined}
}
\item $E_\sigma$ is energy of $\Phi$ with the momentum $\bs P_\sigma$, while $\E_\sigma$ is its complex extension that includes the effect of $\Phi$-decay width in the imaginary part:
\al{
E_\sigma
	&:=	(M^2+\bs P_\sigma^2)^{1/2},&
\E_\sigma
	&:=	\sqrt{E_\sigma^2-i\epsilon}
	\simeq
		E_\sigma-{i\epsilon\ov 2E_\sigma}.
		\label{E_sigma defined}
}
\end{itemize}

At the leading order in the plane-wave expansion for large $\sigma_a$, the Gaussian wave packet~\eqref{full Gaussian} is approximated by~\cite{Ishikawa:2005zc,Ishikawa:2018koj,Ishikawa:2020hph}
\al{
f_{\sigma_a;X_a,\bs P_a}\fn{z}
	&\simeq
			{e^{iP_a\cdot\pn{z-X_a}}\ov\pn{\pi\sigma_a}^{3/4}\sqrt{2E_a}}
			e^{-{1\ov2\sigma_a}\sqbr{\bs z-\bs X_a-\pn{z^0-X_a^0}\bs V_a}^2}.
			\label{large sigma for f}
}
After putting them into Eq.~\eqref{original S in Supplemental Material}, we can perform the six-dimensional Gaussian integral over the spatial positions of in- and out-interaction vertices $\bs x$ and $\bs y$, respectively, without any further approximation.
The resultant wave-packet S-matrix becomes
\al{
\mc S
	&\simeq	
		\Pn{2\pi\sqrt{\sigma_\tx{in}\sigma_\tx{out}}}^3
		N
		\int{\df^3\bs p\ov\pn{2\pi}^3}e^{
			-{\sigma_\tx{in}\ov2}\paren{\bs p-\bs P_\tx{in}}^2
			-{\sigma_\tx{out}\ov2}\paren{\bs p-\bs P_\tx{out}}^2
			-i\ol{\bs\Xi}_\tx{in}\cdot\paren{\bs p-\bs P_\tx{in}}
			+i\ol{\bs\Xi}_\tx{out}\cdot\paren{\bs p-\bs P_\tx{out}}
			}\nn
	&\quad\times
		\int_{-\infty}^\infty\df x^0
					e^{
			-{1\ov2\stin}\paren{x^0-\mf T_\tx{in-int}}^2
			-i\omega_\tx{in}\fn{\bs p}x^0
			}
		\int_{-\infty}^\infty\df y^0e^{
			-{1\ov2\stout}\paren{y^0-\mf T_\tx{out-int}}^2
			+i\omega_\tx{out}\fn{\bs p}y^0
			}\nn
	&\quad\times
		\int{\df p^0\ov2\pi}
			{-i\,e^{-ip^0\pn{y^0-x^0}}\ov p^2+M^2-i\epsilon},
			\label{original S}
}
where we have defined the following:
\begin{itemize}
\item $\ol{\bs\Xi}_\tx{in}$ is a weighted average of the center positions of the two incoming wave packets $\phi\phi$ at an arbitrary reference time $t=0$,
\al{
\ol{\bs\Xi}_\tx{in}
	&:=	{\sigma_1\sigma_2\ov\sigma_1+\sigma_2}\pn{{\bs\Xi_1\ov\sigma_1}+{\bs\Xi_2\ov\sigma_2}},
}
and similarly $\ol{\bs\Xi}_\tx{out}$ is of the two outgoing ones,
\al{
\ol{\bs\Xi}_\tx{out}
	&:=	{\sigma_3\sigma_4\ov\sigma_3+\sigma_4}\pn{{\bs\Xi_3\ov\sigma_3}+{\bs\Xi_4\ov\sigma_4}},
}
where the center of $a$th wave packet at $t=0$ is defined by\footnote{
$\bs\Xi_a$ is $\left.\bs\Xi_a\fn{t}\right|_{t=0}=:\boldsymbol{\mathfrak X}_a$ in the language of Ref.~\cite{Ishikawa:2020hph}.
}
\al{
\bs\Xi_a
	&:=	\bs X_a-\bs V_aX_a^0;
}
see Fig.~\ref{schematic figure}.
\item
$\mf T_\tx{in-int}$ ($\mf T_\tx{out-int}$) is the intersection time for the incoming (outgoing) $\phi\phi$:
\al{
\mf T_\tx{in-int}
	&=	-{\pn{\bs V_1-\bs V_2}\cdot\pn{\bs\Xi_1-\bs\Xi_2}\ov\pn{\bs V_1-\bs V_2}^2},&
\mf T_\tx{out-int}
	&=	-{\pn{\bs V_3-\bs V_4}\cdot\pn{\bs\Xi_3-\bs\Xi_4}\ov\pn{\bs V_3-\bs V_4}^2};
}
see Fig.~\ref{schematic figure}.
\item
$x^0$ ($y^0$) is the time for the in(out)-interaction vertex.
%
\item
$\omega_\tx{in}\fn{\bs p}$ ($\omega_\tx{out}\fn{\bs p}$) is the ``shifted energy'' for the in (out) interaction:
\al{
\omega_\tx{in}\fn{\bs p}
	&:=	E_\tx{in}+\ol{\bs V}_\tx{in}\cdot\pn{\bs p-\bs P_\tx{in}},\nn
\omega_\tx{out}\fn{\bs p}
	&:=	E_\tx{out}+\ol{\bs V}_\tx{out}\cdot\pn{\bs p-\bs P_\tx{out}}.
	\label{omega defined}
}
\item $N$ is a normalization factor that is mostly irrelevant to the following discussion:
\al{
N:=\pn{-i\kappa}^2
		e^{-{\mc R_\tx{in}+\mc R_\tx{out}\ov2}}
			\prod_{a=1}^4{1\ov\sqrt{2E_a}}
			\pn{1\ov\pi\sigma_a}^{3/4},
			\label{N factor}
}
in which the ``overlap exponents'' are given by\footnote{
The wave limit of $\mc R$, Eq.~(C.5), in Ref.~\cite{Ishikawa:2018koj} has a typo in the sign of the second term.
}
\al{
\mc R_\tx{in}
	&=	{
		\pn{\bs\Xi_1-\bs\Xi_2}^2
		-\sqbr{\wh{\bs V}_{12}\cdot\pn{\bs\Xi_1-\bs\Xi_2}}^2
			\ov\sigma_1+\sigma_2},&
\mc R_\tx{out}
	&=	{
		\pn{\bs\Xi_3-\bs\Xi_4}^2
		-\sqbr{\wh{\bs V}_{34}\cdot\pn{\bs\Xi_3-\bs\Xi_4}}^2
			\ov\sigma_3+\sigma_4},
}
with
$\wh{\bs V}_{12}
	:=	{\bs V_1-\bs V_2\ov\ab{\bs V_1-\bs V_2}}$ and
$\wh{\bs V}_{34}
	:=	{\bs V_3-\bs V_4\ov\ab{\bs V_3-\bs V_4}}$.
\end{itemize}
Here and hereafter, we neglect the overall phase factor that is independent of the integrated variables $\pn{p^0,\bs p,x^0,y^0}$.

\subsection{Strategy of paper}
In this paper, we evaluate the S-matrix~\eqref{original S} in two different orders of integrals:
\begin{enumerate}[Order 1.]
\item First integrate over $\bs p$, $p^0$, and $x^0$. Then we exhibit the \emph{emergence} of an in-time-boundary effect for the $\Phi\to\phi\phi$ decay from the remaining integral over $y^0$.\label{order 1}
\item First integrate over $x^0$ and $y^0$. Then we analyze the structure of the off-shell $p^0$ integral via two distinct approaches:\label{order 2}
\begin{enumerate}[(a)]
\item The saddle-point method and the residue theorem\label{approach a}
\item The Lefschetz thimble decomposition\label{approach b}
\end{enumerate}
\end{enumerate}
We will also show that the result of the remaining $\bs p$ integral in a non-relativistic limit in Order~\ref{order 2} agrees with that of Order~\ref{order 1}.
Roughly, Order~\ref{order 1} examines the emergence of the boundary for $y^0$ (the time of local interaction governing $\Phi\to\phi\phi$) in the position space, whereas Order~\ref{order 2} examines the structure of the $\Phi$ propagator in its momentum space.

Finally, we comment on the possible ``time-boundary effect'' of $\phi$. 
In this paper, we first \emph{neglect} the time boundaries of $\phi$
and then will show that the in-time-boundary effect of $\Phi$ for the $\Phi\to\phi\phi$ decay still emerges.\footnote{
This may be rephrased as follows:
We first introduce time boundaries for the two-to-two scattering, $T_\tx{in}$ and $T_\tx{out}$, at which interactions are negligible; cut off the interaction-time integrals $\int_{T_\tx{in}}^{T_\tx{out}}\df x^0$ and $\int_{T_\tx{in}}^{T_\tx{out}}\df y^0$; focus on the ``bulk terms'' for the two-to-two scattering; and take the limit $T_\tx{in}\to-\infty$ and $T_\tx{out}\to\infty$.
}

\section{Emergence of time boundary for $\Phi\to\phi\phi$}\label{emergence of time boundary section}
First, we run a series of calculations in Sec.~\ref{calculation subsection} through Order~\ref{order 1} from the S-matrix~\eqref{original S}.
Then we will discuss its properties in Sec.~\ref{properties subsection}.

\subsection{Calculation of Gaussian S-matrix}\label{calculation subsection}
From the S-matrix~\eqref{original S}, we perform the Gaussian integral over $\bs p$ in the saddle-point approximation for large $\sigma_+$. The result is
\al{
\mc S
	&\simeq
		\paren{2\pi\sigma}^{3/2}N
		\int_{-\infty}^\infty\df x^0\,e^{
			-{1\ov2\stin}\paren{x^0-\mf T_\tx{in-int}}^2
			}
		\int_{-\infty}^\infty\df y^0\,e^{
			-{1\ov2\stout}\paren{y^0-\mf T_\tx{out-int}}^2
			}
		e^{V_\fs\fn{x^0,y^0}}
		\mc I_\fs\fn{x^0,y^0},
		\label{result of momentum integral}
}
where\footnote{
Starting from Eq.~\eqref{original S}, we have an extra phase factor $e^{-i\ol{\bs\Xi}_\sigma\cdot\pn{\bs P_\tx{out}-\bs P_\tx{in}}}$ with
$\ol{\bs\Xi}_\sigma
	:=	{\sigma_\tx{in}\ol{\bs\Xi}_\tx{in}+\sigma_\tx{out}\ol{\bs\Xi}_\tx{out}\ov\sigma_\tx{in}+\sigma_\tx{out}}$. As said above, we neglect such a phase factor that does not contain any integration variable.
}
\al{
\mc I_\fs\fn{x^0,y^0}
	&:=	\int{\df p^0\ov2\pi i}
			{e^{-ip^0\pn{y^0-x^0}}\ov-\pn{p^0}^2+\bs p_\fs^2+M^2-i\epsilon},
					\label{I star}\\
V_\fs\fn{x^0,y^0}
	&:=
		-{\sigma\ov2}\pn{\bs P_\tx{out}-\bs P_\tx{in}}^2
		-{\pn{
			\delta\ol{\bs\Xi}+\ol{\bs V}_\tx{out}y^0
			-\ol{\bs V}_\tx{in}x^0}^2\ov2\sigma_+}\nn
	&\quad
		-i\sigma\pn{
				{\ol{\bs V}_\tx{in}\ov\sigma_\tx{in}}x^0
				+{\ol{\bs V}_\tx{out}\ov\sigma_\tx{out}}y^0
				}
			\cdot\pn{\bs P_\tx{out}-\bs P_\tx{in}}
		-i\omega_\tx{in}\fn{0}x^0+i\omega_\tx{out}\fn{0}y^0,
			\label{V star}
}
in which
\al{
\delta\ol{\bs\Xi}
	:=	\ol{\bs\Xi}_\tx{out}-\ol{\bs\Xi}_\tx{in}
		\label{delta Xi bar}
}
is the displacement between the in and out wave packets and the saddle point is located at
\al{
\bs p_\fs
	&:=	
		\bs P_\sigma+i{	
			\delta\ol{\bs\Xi}+\ol{\bs V}_\tx{out}y^0-\ol{\bs V}_\tx{in}x^0
			\ov\sigma_+}.
}
Recall that $\sigma_+$ and $\bs P_\sigma$ have been defined in Eqs.~\eqref{sigma plus} and \eqref{P_sigma defined}, respectively.
The mass dimensions of $\mc I_\fs$ and $V_\fs$ are $-1$ and 0, respectively.


Hereafter, we replace $\bs p_\fs$ by $\bs P_\sigma$ in Eq.~\eqref{I star}, namely, we take the leading-order terms in the large $\sigma_+$ expansion in its polynomial while leaving the exponents intact:
\al{
\mc I_\fs\fn{x^0,y^0}
	&\simeq
		\int{\df p^0\ov2\pi i}
			{e^{-ip^0\pn{y^0-x^0}}\ov
				-\pn{p^0}^2+\bs P_\sigma^2
				+M^2
				-i\epsilon
				}.
				\label{expanded expression}
}
Then we can perform the $p^0$ integral analytically:
\al{
\mc I_\fs\fn{x^0,y^0}
	&=
		{1\ov2\E_\sigma}
			\pn{
				\theta\fn{y^0-x^0}e^{-i\E_\sigma\pn{y^0-x^0}}
				+\theta\fn{x^0-y^0}e^{i\E_\sigma\pn{y^0-x^0}}
				},
				\label{I star approximated}
}
where $\theta$ is the Heaviside step function. Recall that $\E_\sigma$ is defined in Eq.~\eqref{E_sigma defined}.
For the propagation from the in-interaction time $x^0$ to the out-interaction time $y^0$, the first and second terms correspond to the forward and backward propagation in time, respectively.

After putting the result~\eqref{I star approximated}, we may perform the $x^0$ integral in Eq.~\eqref{result of momentum integral} exactly. The result is
\al{
\mc S
	&=	{\paren{2\pi\sigma}^{3/2}N\ov2\E_\sigma}
		e^{
			-{\sigma\ov2}\pn{\bs P_\tx{out}-\bs P_\tx{in}}^2
			-{1\ov2\sigma_+}\pn{\ol{\bs\Xi}_\tx{out}-\ol{\bs\Xi}_\tx{in}}^2
			-{\mf T_\tx{in-int}^2\ov2\stin}
			-{\mf T_\tx{out-int}^2\ov2\stout}
			}
		\sqrt{2\pi\s_\tx{in}}\int_{-\infty}^\infty\df y^0\nn
	&\quad\times
		\Bigg[
			e^{
			-{\pn{y^0-\T_\tx{out-int}^+}^2\ov\s_\tx{out}}
			+{\pn{\T_\tx{in-int}^+}^2\ov2\s_\tx{in}}
			+{\pn{\T_\tx{out-int}^+}^2\ov2\s_\tx{out}}
			}
			{1
			+\erf\fn{
				\pn{1-{\s_\tx{in}\ov\sigma_+}\ol{\bs V}_\tx{in}\cdot\ol{\bs V}_\tx{out}}y^0
				-\T_\tx{in-int}^+
					\ov\sqrt{2\s_\tx{in}}
				}\ov2}
				\nn
	&\phantom{\quad\times\Bigg[}
			+e^{
			-{\pn{y^0-\T_\tx{out-int}^-}^2\ov\s_\tx{out}}
			+{\pn{\T_\tx{in-int}^-}^2\ov2\s_\tx{in}}
			+{\pn{\T_\tx{out-int}^-}^2\ov2\s_\tx{out}}
			}
			{1+\erf\fn{
				\T_\tx{in-int}^-
				-\pn{1-{\s_\tx{in}\ov\sigma_+}\ol{\bs V}_\tx{in}\cdot\ol{\bs V}_\tx{out}}y^0
					\ov\sqrt{2\s_\tx{in}}
				}\ov2}
		\Bigg],\label{y0 left}
}
where $\s_\tx{in}$ and $\s_\tx{out}$ are effective temporal width-squared for the in- and out-intersections, respectively,
\al{
\s_\tx{in}
	&:=	{1\ov{1\ov\stin}+{\ol{\bs V}_\tx{in}^2\ov\sigma_+}},\\
\s_\tx{out}
	&:=	{\stout\ov1+{\stout\ov\sigma_+}
			\pn{\ol{\bs V}_\tx{out}^2
				-{\s_\tx{in}\pn{\ol{\bs V}_\tx{in}\cdot\ol{\bs V}_\tx{out}}^2\ov\sigma_+}}},
}
and the complex parameters $\T_\tx{in-int}$ and $\T_\tx{out-int}$ correspond to the in- and out-intersection times $\mf T_\tx{in-int}$ and $\mf T_\tx{out-int}$ (see Fig.~\ref{schematic figure}), respectively,
\al{
\T_\tx{in-int}^\pm
	&:=	
		\s_\tx{in}\br{
			\sqbr{{\mf T_\tx{in-int}\ov\stin}
				+{\ol{\bs V}_\tx{in}\cdot\delta\ol{\bs\Xi}\ov\sigma_+}
				}
			-i\sqbr{\omega_\tx{in}\fn{0}
				\mp\E_\sigma
				+{\sigma\ol{\bs V}_\tx{in}\cdot\pn{\bs P_\tx{out}-\bs P_\tx{in}}\ov\sigma_\tx{in}}
					}
			},\\
\T_\tx{out-int}^\pm
	&:=	\s_\tx{out}\Bigg\{
			\sqbr{
				{\mf T_\tx{out-int}\ov\stout}
				+{\ol{\bs V}_\tx{out}\cdot\pn{
						\ol{\bs V}_\tx{in}\T_\tx{in-int}
						-\delta\ol{\bs\Xi}}
					\ov\sigma_+}
				}\nn
	&\phantom{:=\s_\tx{out}\Bigg\{}
			+i\sqbr{
				\omega_\tx{out}\fn{0}
				\mp\E_\sigma
					\pn{1-{\s_\tx{in}\ol{\bs V}_\tx{in}\cdot\ol{\bs V}_\tx{out}\ov\sigma_+}	
						}
				-{\sigma\ol{\bs V}_\tx{out}\cdot\pn{\bs P_\tx{out}-\bs P_\tx{in}}\ov\sigma_\tx{out}}
				}
			\Bigg\}.
}

In Eq.~\eqref{y0 left}, we can always separate the following expression for a general complex number~$z$ into what we call the bulk and boundary terms:
\al{
{1+\erf\fn{z\ov\sqrt{2\s_\tx{in}}}\ov2}
	&=	{1+\sgn\fn{z\ov\sqrt{2\s_\tx{in}}}\ov2}+{\erf\fn{z\ov\sqrt{2\s_\tx{in}}}-\sgn\fn{z\ov\sqrt{2\s_\tx{in}}}\ov2}\nn
	&=	\theta\fn{\Re z}+{\erf\fn{z\ov\sqrt{2\s_\tx{in}}}-\sgn\fn{z\ov\sqrt{2\s_\tx{in}}}\ov2},
	\label{erf into theta and boundary}
}
where $\Re$ and $\Im$ denote the real and imaginary parts, respectively, and the last equality holds except at $\Re z=0$, which is out of our interest.\footnote{
The sign function $\sgn\fn{z}$ for a general complex argument $z$ is defined in Eq.~(63) in Ref.~\cite{Ishikawa:2018koj}.
}

For simplicity, we hereafter focus on the limit
\al{
\s_\tx{in}\to0,
\label{sin to zero limit}
}
which picks up only the bulk terms in Eq.~\eqref{y0 left}:
\al{
{1+\erf\fn{z\ov\sqrt{2\s_\tx{in}}}\ov2}
	&\to \theta\fn{\Re z}.
	\label{small s_in limit}
}
Comparing with Eq.~\eqref{I star approximated}, we see that the first and second terms in the square brackets in Eq.~\eqref{y0 left} correspond to the forward and backward propagation from the (effective) in-intersection time $\T_\tx{in-int}^\pm$ to the out-interaction time $y^0$.

Next, we analytically perform the $y^0$ integral in Eq.~\eqref{y0 left} after taking the limit~\eqref{small s_in limit}, namely, $\sigma_+\gg\s_\tx{in}\ol{\bs V}_\tx{in}\cdot\ol{\bs V}_\tx{out}>0$.
The integral of $y^0$ regenerates the combinations~\eqref{erf into theta and boundary} but with different arguments:
\al{
\mc S
	&=
		{\paren{2\pi\sigma}^{3/2}N\ov2\E_\sigma}
		e^{
			-{\sigma\ov2}\pn{\bs P_\tx{out}-\bs P_\tx{in}}^2
			-{1\ov2\sigma_+}\pn{\ol{\bs\Xi}_\tx{out}-\ol{\bs\Xi}_\tx{in}}^2
			-{\mf T_\tx{in-int}^2\ov2\stin}
			-{\mf T_\tx{out-int}^2\ov2\stout}
			}
		\sqrt{2\pi\s_\tx{in}}\sqrt{2\pi\s_\tx{out}}\nn
	&\quad\times\Bigg\{
		e^{
			{\pn{\T_\tx{in-int}^+}^2\ov2\s_\tx{in}}
			+{\pn{\T_\tx{out-int}^+}^2\ov2\s_\tx{out}}
			}
			{1+\erf\fn{
				\T_\tx{out-int}^+
				-\Re\T_\tx{in-int}^+\ov\sqrt{2\s_\tx{out}}
				}\ov2}\nn
	&\phantom{\quad\times\Bigg\{}
		+e^{
			{\pn{\T_\tx{in-int}^-}^2\ov2\s_\tx{in}}
			+{\pn{\T_\tx{out-int}^-}^2\ov2\s_\tx{out}}
			}
		{1+\erf\fn{
			\Re\T_\tx{in-int}^-
			-\T_\tx{out-int}^-\ov\sqrt{2\s_\tx{out}}}\ov2}
		\Bigg\}.
		\label{final full expression}
}
Recall that $\sigma$ and $\vs$ are defined in Eqs.~\eqref{sigma plus} and \eqref{varsigma defined}, respectively.

Again, we can decompose the parts of the form ${1+\erf\pn{\cdots}\ov2}$ into the bulk and boundary terms via Eq.~\eqref{erf into theta and boundary}.
In particular, the resultant step functions in the braces in Eq.~\eqref{final full expression} are
\al{
\theta\fn{
		\Re\T_\tx{out-int}^+
		-\Re\T_\tx{in-int}^+
		},
			\label{theta function first appeared}
}
and
\al{
\theta\fn{
		{\Re\T_\tx{in-int}^-
			-\Re\T_\tx{out-int}^-
			}},
		\label{theta functions}
}
respectively. We see more transparently that the former and latter represent the forward and backward propagation in time. 
Hereafter, we neglect the latter since we are interested in the near-on-shell scattering around the resonance that suppresses the latter exponentially:
\al{
\mc S
	&=
		{\paren{2\pi\sigma}^{3/2}N\ov2\E_\sigma}
		e^{
			-{\sigma\ov2}\pn{\bs P_\tx{out}-\bs P_\tx{in}}^2
			-{1\ov2\sigma_+}\pn{\ol{\bs\Xi}_\tx{out}-\ol{\bs\Xi}_\tx{in}}^2
			-{\mf T_\tx{in-int}^2\ov2\stin}
			-{\mf T_\tx{out-int}^2\ov2\stout}
			}
		\sqrt{2\pi\s_\tx{in}}\sqrt{2\pi\s_\tx{out}}\nn
	&\quad\times\Bigg\{
		e^{
			{\pn{\T_\tx{in-int}^+}^2\ov2\s_\tx{in}}
			+{\pn{\T_\tx{out-int}^+}^2\ov2\s_\tx{out}}
			}
			{1+\erf\fn{
				\T_\tx{out-int}^+
				-\Re\T_\tx{in-int}^+\ov\sqrt{2\s_\tx{out}}
				}\ov2}
		\Bigg\}.
		\label{without backward propagation}
}

\subsubsection{Simple configuration}
The physical meaning of the general result~\eqref{without backward propagation} is more apparent when we concentrate on the particularly simple configuration:\footnote{
For the simple configuration~\eqref{simple configuration}, we do not need the expansion~\eqref{expanded expression} to proceed with the subsequent computation.
}
\al{
\ol{\bs V}_\tx{in}
	&\to	0,&
\ol{\bs V}_\tx{out}
	&\to	0.
\label{simple configuration}
}
In this case, we get
\al{
\omega_\tx{in}\fn{\bs p}
	&\to E_\tx{in},&
\s_\tx{in}
	&\to	\stin\quad(\sr{\tx{Eq.\,\eqref{sin to zero limit}}}\to0),&
\T_\tx{in-int}
	&\to
		\mf T_\tx{in-int}-i\stin E_\tx{in}
		\quad(\sr{\tx{Eq.\,\eqref{sin to zero limit}}}\to\mf T_\tx{in-int}),\\
\omega_\tx{out}\fn{\bs p}
	&\to E_\tx{out},&
\s_\tx{out}
	&\to	\stout,&
\T_\tx{out-int}
	&\to
		\mf T_\tx{out-int}+i\stout E_\tx{out}.
}
We also note that $\vs\to\stin$ ($\to0$) and $\vs_+\to\stout$ due to Eq.~\eqref{sin to zero limit}.
The following simplification will also be useful for the quantities defined below in Eqs.~\eqref{omega vs defined} and \eqref{delta omega defined}:
\al{
\delta\omega\fn{\bs p}
	&\to
		E_\tx{out}-E_\tx{in},&
\omega_\vs\fn{0}
	&\to
		E_\vs\quad(\sr{\tx{Eq.\,\eqref{sin to zero limit}}}\to E_\tx{out}),
}
where
\al{
E_\vs
	&:=	{\vs_\tx{in}E_\tx{in}+\vs_\tx{out}E_\tx{out}\ov\vs_\tx{in}+\vs_\tx{out}}
		\quad(\sr{\tx{Eq.\,\eqref{sin to zero limit}}}\to E_\tx{out}).
}

The result~\eqref{without backward propagation} simplifies to
\al{
\mc S
	&\to
		\pn{2\pi}^4N
		\sqbr{
			\paren{\sigma\ov2\pi}^{3\ov2}
			e^{-{\sigma\ov2}\pn{\bs P_\tx{out}-\bs P_\tx{in}}^2}
			}
		\sqbr{
			\sqrt{\stin\ov2\pi}
			e^{-{\stin\ov2}\pn{E_\tx{out}-E_\tx{in}}^2}
			}
		{\sqrt{2\pi\stout}\ov2\E_\sigma}
		e^{	-{\stout\ov2}\pn{\E_\sigma-E_\tx{out}}^2
			-{1\ov2\sigma_+}\pn{\ol{\bs\Xi}_\tx{out}-\ol{\bs\Xi}_\tx{in}}^2}\nn
	&\quad\times
		\br{
			\theta\fn{\Re{\DelT\ov\sqrt{2\stout}}}
			+
			{\erf\fn{{\DelT\ov\sqrt{2\stout}}}
				-\sgn\fn{{\DelT\ov\sqrt{2\stout}}}\ov2}
		},
%
		\label{emerging 1 to 2}
}
where
\al{
\delta\mf T
	&:=	\mf T_\tx{out-int}-\mf T_\tx{in-int},&
\DelT
	&:=	\delta\mf T+\stout\Im\E_\sigma+i\stout\pn{E_\tx{out}-\Re\E_\sigma}.
}
Here, $\delta\mf T$ is the elapsed time between the in and out intersections, namely the propagation time of $\Phi$; see Fig.~\ref{schematic figure}.
The physical meaning of each factor in Eq.~\eqref{emerging 1 to 2} is the following:
\begin{itemize}
\item $N$ is the factor~\eqref{N factor} that depends on the configuration of external states. This factor is out of our interest in this paper.
\item The first and second square brackets represent the momentum and energy conservations, and reduce to the delta functions $\delta^3\fn{\bs P_\tx{out}-\bs P_\tx{in}}$ and $\delta\fn{E_\tx{out}-E_\tx{in}}$ in the limits of $\sigma\to\infty$ and $\vs_\tx{in}\to\infty$, respectively.
\item The complex parameter $\E_\sigma$ is the energy~\eqref{E_sigma defined} of the intermediate particle $\Phi$ that includes its width as well.
\item Eq.~\eqref{emerging 1 to 2} contains the $\Phi\to\phi\phi$ amplitude, Eq.~(58) in Ref.~\cite{Ishikawa:2018koj}.\footnote{
$E_\sigma$ is identified to its initial energy $E_0$,
$\sigma_\tx{out}$ to its width-squared $\sigma_s$,
$\stout$ to its temporal width-squared $\sigma_t$,
$\mf T_\tx{in-int}$ to its in time boundary $T_\tx{in}$, and 
$\mf T_\tx{out-int}$ to its interaction time $\mf T$.
Note that $\sigma_\tx{in}$ and $\stin$ have no counterparts in Ref.~\cite{Ishikawa:2018koj} and that the $\phi\phi$ out time boundary has been neglected in this paper.
}
The first and second terms in the braces in Eq.~\eqref{emerging 1 to 2} are the bulk and in-time-boundary terms, respectively, introduced in Eq.~\eqref{erf into theta and boundary}.
\end{itemize}
For the first time, we have \emph{proven} that the in time boundary of the $\Phi\to\phi\phi$ decay \emph{emerges} even if we do not take into account \emph{any} time-boundary effect for the $\phi\phi\to\phi\phi$ scattering.

\subsection{Properties of Gaussian S-matrix}\label{properties subsection}
Now we examine various detailed aspects of the Gaussian wave-packet S-matrix~\eqref{emerging 1 to 2}, obtained for the simple configuration~\eqref{simple configuration}, by varying the configuration of external states.
%
%
%
%


\begin{figure}\centering
\includegraphics[width=0.8\textwidth]{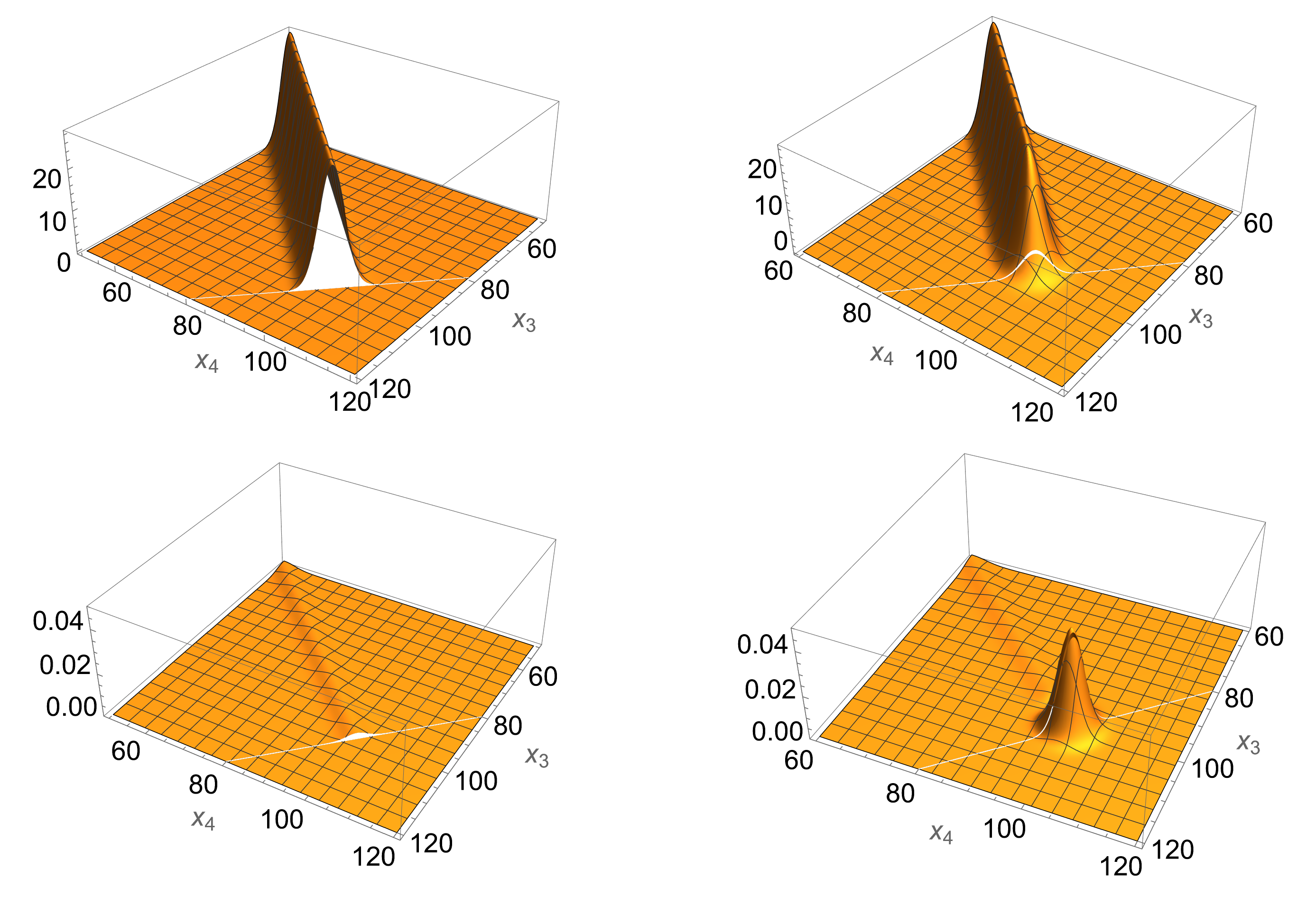}
\caption{Differential probability $\ab{\mc S}^2/{\kappa^4\ov64\pi M^6}{\sigma_1\ov\sigma_3^2}$ from Eq.~\eqref{emerging 1 to 2} as a function of the final wave-packet positions $x_3:=\ab{\bs X_3}$ and $x_4:=\ab{\bs X_4}$ at time $X_3^0=X_4^0=:T$ ($=100$ in units $M=1$). The upper and lower panels are for momenta on and off the resonance, respectively. 
The left panels show the bulk contribution only, whereas the right ones show the sum of all the bulk and boundary ones.
The difference between the right and left panels is the boundary contribution.
This figure verifies our theoretical prediction that the boundary effect is sizable both on and off the resonance.
\label{illustration}}
\end{figure}

As an illustration, in Fig.~\ref{illustration}, we show the dependences of the (normalized) differential probability
\al{
\ab{\mc S}^2\bigg/{\kappa^4\ov64\pi M^6}{\sigma_1\ov\sigma_3^2}
}
on the final wave-packet positions $x_3:=\ab{\bs X_3}$ and $x_4:=\ab{\bs X_4}$ at time $X_3^0=X_4^0=T$ for a typical setup closest to the on-shell plane-wave calculation. The position vectors of initial wave packets are back to back and collinear to their momenta, corresponding to a head-on collision, with the collision point being the origin of the spacetime ($\mf T_\tx{in-int}=0$), and the final state momenta are back to back as well (we focus on the case of symmetric widths $\sigma_1=\sigma_2\ll\sigma_3=\sigma_4$ and the on-shell momenta $E_1+E_2=E_3+E_4$ for a small~$\epsilon$). Then we obtain $\delta\ol{\bs\Xi}
={x_3-x_4\ov2}\bs V_3$ ($=\ol{\bs\Xi}_\tx{out}$) and $\delta\mf T=T-{x_3+x_4\ov2}$ ($=\mf T_\tx{out-int}$). $\delta\mf T>0$ and $\delta\mf T=0$ correspond to the bulk and in time boundary for the $\Phi\to\phi\phi$ decay.\footnote{
Here and in the next paragraph, (in)equalities are given for $\epsilon\to0$ for simplicity. If we recover it, the bulk and in time boundary are $\delta\mf T>{\epsilon\sigma_3\ov4M\bs V_3^2}$ and $\delta\mf T={\epsilon\sigma_3\ov4M\bs V_3^2}$, respectively.
}
Further details are presented in Appendix~\ref{Details on Fig 2}.

We first discuss the left panels that include the bulk contribution at ${x_3+x_4\ov2}<T$, with ${x_3+x_4\ov2}= T$ being the in time boundary.
The probability density $\ab{\mc S}^2$ is 
constant along the $x_3=x_4$ line, exhibiting manifest independence on $\delta\mf T$.
Along the perpendicular direction to this constant line, $\ab{\mc S}^2$ is exponentially damped with the width $\sim \sqrt{\sigma_3}/\ab{\bs V_3}$. 
This width increases as the size of the wave packets $\sqrt{\sigma_3}$ increases, but the in time boundary stays at ${x_3+x_4\ov2}= T$.
$\ab{\mc S}^2$ summed over $x_3$ and $x_4$ remains trivially independent of $\delta\mf T$.

Next we discuss the right panels with all the contributions.
$\ab{\mc S}^2$ is smooth everywhere thanks to the boundary contribution localized around $x_3\simeq x_4\simeq T$ ($=100$), namely $\delta\mf T\simeq0$:
\begin{itemize}
\item On the resonance (in the upper-right panel), the interference between the bulk and boundary terms provides a negative contribution at $\delta\mf T\gtrsim0$ ($x_3+x_4\lesssim T$), while the absolute-square of the boundary term gives a sizable contribution at $\delta\mf T\lesssim 0$ ($x_3+x_4\gtrsim T$).
\item Off the resonance (in the lower-right panel), the boundary effect is prominent. The boundary contribution is suppressed only by a power law at the off-resonance, while the bulk one is by the exponential law.
\end{itemize}
The large boundary effect at off-resonance has been overlooked because it is buried in backgrounds or has itself been regarded as a background. Its confirmation requires a dedicated experiment or a comprehensive reanalysis of existing data~\cite{Ishikawa:2019nes,Ushioda:2019hje}.

The wave-packet formalism contains more complete information than the plane-wave formalism.
For the absolute-square of the S-matrix~\eqref{emerging 1 to 2}, the phase space of the initial and final states are spanned not only by $\bs P_1,\dots,\bs P_4$ but also $\bs X_1,\dots,\bs X_4$. The latter information is peculiar to the wave-packet formalism, and is encoded only in $\delta\ol{\bs\Xi}$ and $\delta\mf T$ (which also appears in $\Delta\T$), always appearing as the ratios $\delta\ol{\bs\Xi}/\sqrt{\sigma_+}$ and $\delta\mf T/\sqrt{\vs_\tx{out}}$, respectively.
When $\ab{\delta\ol{\bs\Xi}}\lesssim\sqrt{\sigma_+}$, the wave packets sizably overlap with each other, and the probability $|\mc S|^2$ is not suppressed by the overall factor $e^{-(\delta\ol{\bs\Xi})^2/\sigma_+}$.
Among such configurations,
when $|\delta\mf T|\gg\sqrt{\vs_\tx{out}}$ the boundary term is exponetially suppressed and the bulk term dominates, while when $|\delta\mf T|\lesssim\sqrt{\vs_\tx{out}}$, the boundary contribution becomes sizable and cannot be neglected.
We note that $\delta\mf T$ takes all the values $-\infty<\delta\mf T<\infty$ as we vary $\bs X_1,\dots,\bs X_4$.

Finally, we show the result of a limit $\ab{\DelT}\gg\sqrt{2\vs_\tx{out}}$ with $\delta\mf T>0$ that makes the analytic structure of the amplitude~\eqref{emerging 1 to 2} more apparent:
\al{
\mc S
	&\to
		\pn{2\pi}^4N
		\sqbr{\paren{\sigma\ov2\pi}^{3/2}
		e^{
			-{\sigma\ov2}\pn{\bs P_\tx{out}-\bs P_\tx{in}}^2
		}}
		\sqbr{
		\sqrt{\stin\ov2\pi}
		e^{
			-{\stin\ov2}\pn{E_\tx{out}-E_\tx{in}}^2
			}}
		e^{-{\pn{\ol{\bs\Xi}_\tx{out}-\ol{\bs\Xi}_\tx{in}}^2\ov2\sigma_+}}
		{\sqrt{2\pi\stout}\ov2\E_\sigma}\nn
	&\quad\times
		\Bigg\{
			\theta\fn{\delta\mf T+\stout\Im\E_\sigma}
				e^{-{\stout\ov2}\pn{E_\tx{out}-\E_\sigma}^2}
				-{1\ov\sqrt{2\pi\stout}}{-ie^{-{\pn{\delta\mf T+\stout\Im\E_\sigma}^2\ov2\stout}-i\pn{E_\tx{out}-E_\sigma}\pn{\delta\mf T+\stout\Im\E_\sigma}}\ov
			E_\tx{out}-\E_\sigma-i{\delta\mf T\ov\stout}
			}
		\Bigg\}.
		\label{limiting behavior of boundary and bulk}
}
This form will be used for later comparison.

\section{Alternative derivations via Order 2}\label{Order 2 section}
Now we turn to Order~\ref{order 2}.
Originally, the exponents in Eq.~\eqref{original S} are linear in $p^0$.
We perform $x^0$ and $y^0$ integrals exactly in Eq.~\eqref{original S},
\al{
\mc S
	&=	\pn{2\pi}^4
			\sqrt{\sigma_\tx{in}^3\sigma_\tx{out}^3\vs_\tx{in}\vs_\tx{out}}
		N
		\int{\df^4p\ov\pn{2\pi}^4}{-ie^{f\pn{p^0;\bs p}}\ov p^2+M^2-i\epsilon},
		\label{S-matrix after 8 integrals}
}
and the exponent becomes quadratic in $p^0$:
\al{
f\fn{p^0;\bs p}
	=	f_*\fn{\bs p}-{\vs_+\ov2}\pn{p^0-p^0_*\fn{\bs p}}^2,
}
where the location of saddle point on the complex $p^0$ plane becomes
\al{
p^0_*\fn{\bs p}
	&=	\omega_\vs\fn{\bs p}
		-i{\delta\mf T\ov\vs_+},
		\label{p0star given}
}
in which
\al{
\omega_\vs\fn{\bs p}:={\stin\omega_\tx{in}\fn{\bs p}+\stout\omega_\tx{out}\fn{\bs p}\ov\stin+\stout}
	\label{omega vs defined}
}
is the off-shell shifted energy of $\Phi$.
Here, the exponent at the saddle point becomes
\al{
f_*\fn{\bs p}
	&=	\pn{
			-{\sigma_\tx{in}\ov2}\paren{\bs p-\bs P_\tx{in}}^2
			-i\ol{\bs\Xi}_\tx{in}\cdot\paren{\bs p-\bs P_\tx{in}}
			}
		+\pn{
			-{\sigma_\tx{out}\ov2}\paren{\bs p-\bs P_\tx{out}}^2
			+i\ol{\bs\Xi}_\tx{out}\cdot\paren{\bs p-\bs P_\tx{out}}
			}\nn
	&\quad
		-{\pn{\delta\mf T}^2\ov2\varsigma_+}
		+i\varsigma\pn{
			{\mf T_\tx{in-int}\ov\varsigma_\tx{in}}
			+{\mf T_\tx{out-int}\ov\varsigma_\tx{out}}
			}\delta\omega\fn{\bs p}
		-{\varsigma\ov2}\pn{\delta\omega\fn{\bs p}}^2,
}
where
\al{
\delta\omega\fn{\bs p}
	&:=	\omega_\tx{out}\fn{\bs p}-\omega_\tx{in}\fn{\bs p};
	\label{delta omega defined}
}
see Eq.~(107) in Ref.~\cite{Ishikawa:2020hph}.
That is, the structure of the exponent is
\al{
f_*\fn{\bs p}=-{\pn{\delta\mf T}^2\ov2\vs_+}+\tx{``$\bs p$-dependent terms''}.
}

%
The integral in Eq.~\eqref{S-matrix after 8 integrals}, which we call the ``wave-packet Feynman propagator,'' may be written as
\al{
\int{\df^3\bs p\ov\pn{2\pi}^3}e^{f_*\fn{\bs p}}I\fn{\bs p},
\label{wave-packet Feynman propagator equation}
}
where
\al{
I\fn{\bs p}
	:=	\int_{-\infty}^\infty{\df p^0\ov2\pi}{-i\ov-\pn{p^0}^2+E_{\bs p}^2-i\epsilon}e^{-{\vs_+\ov2}\pn{p^0-p^0_*\pn{\bs p}}^2}
	=	\int_{-\infty}^\infty{\df p^0\ov2\pi}{-i\ov-\pn{p^0}^2+\E_{\bs p}^2}e^{-{\vs_+\ov2}\pn{p^0-p^0_*\pn{\bs p}}^2}.
		\label{I appearing again}
}

\subsection{(a) Saddle point and poles}\label{approach a section}
The above exponential factor has a saddle point at $p^0=p^0_*\fn{\bs p}$, with its steepest descent and ascent paths being $\Im\fn{p^0-p^0_*\fn{\bs p}}=0$ and $\Re\fn{p^0-p^0_*\fn{\bs p}}=0$, respectively. Along the former, the saddle-point approximation gives
\al{
I_*\fn{\bs p}
	\approx{1\ov\sqrt{2\pi\vs_+}}{-i\ov-(\omega_\vs\fn{\bs p}-i{\delta\mf T\ov\vs_+})^2+E_{\bs p}^2-i\epsilon},
}
which leads to Eq.~(110) in Ref.~\cite{Ishikawa:2020hph}.

We compute the wave-packet Feynman propagator:
\al{
\int{\df^4p\ov\pn{2\pi}^4}{-i\ov p^2+M^2-i\epsilon}e^{f\pn{p^0;\bs p}}
	&=	\int{\df^3\bs p\ov\pn{2\pi}^3}e^{f_*\fn{\bs p}}I\fn{\bs p}.
		\label{wave-packet Feynman propagator}
}
An important observation is that the saddle-point integral $I_*$ over the steepest descent path from $-\infty+\Im p^0_*\fn{\bs p}$ to $\infty+\Im p^0_*\fn{\bs p}$ differs from the integral $I$ over $\mathbb R$ by the residues:
\al{
I\fn{\bs p}
	&=	I_*\fn{\bs p}
		-\pn{\underset{{p^0=\E_{\bs p}}}{\Res}
			{e^{-{\vs_+\ov2}\pn{p^0-p^0_*\pn{\bs p}}^2}\ov-\pn{p^0}^2+\E_{\bs p}^2}}
			\theta\fn{\Im\E_{\bs p}-\Im p^0_*}
		+\pn{\underset{{p^0=-\E_{\bs p}}}{\Res}
			{e^{-{\vs_+\ov2}\pn{p^0-p^0_*\pn{\bs p}}^2}\ov-\pn{p^0}^2+\E_{\bs p}^2}}
			\theta\fn{\Im p^0_*-\Im\fn{-\E_{\bs p}}}\nn
	&=
		I_*\fn{\bs p}
		+{e^{-{\vs_+\ov2}\pn{\omega_\vs\pn{\bs p}-\E_{\bs p}-i{\delta\mf T\ov\vs_+}}^2}\ov2\E_{\bs p}}
			\theta\fn{{\delta\mf T\ov\vs_+}+\Im\E_{\bs p}}
		+{e^{-{\vs_+\ov2}\pn{\omega_\vs\pn{\bs p}+\E_{\bs p}-i{\delta\mf T\ov\vs_+}}^2}\ov2\E_{\bs p}}
			\theta\fn{\Im\E_{\bs p}-{\delta\mf T\ov\vs_+}},
			\label{I integral}
}
where
\al{
I_*\fn{\bs p}
	&:=
		\int_{-\infty+\Im p^0_*}^{\infty+\Im p^0_*}{\df p^0\ov2\pi}{-i\ov-\pn{p^0}^2+E_{\bs p}^2-i\epsilon}e^{-{\vs_+\ov2}\pn{p^0-p^0_*\pn{\bs p}}^2}
	\approx
		{1\ov\sqrt{2\pi\vs_+}}{-i\ov-(\omega_\vs\fn{\bs p}-i{\delta\mf T\ov\vs_+})^2+E_{\bs p}^2-i\epsilon}.
			\label{I star integral}
}
The exponent of the first term in Eq.~\eqref{I integral} is quadratic in $\bs p$ after the $p^0$-integral~\eqref{I star integral} and can be computed as in Eq.~(134) in Ref.~\cite{Ishikawa:2020hph},
while that of the second term is not, and we compute it in the non-relativistic limit below.
We will neglect the third term, which corresponds to a propagation backward in time, assuming a nearly on-shell scattering.

\begin{figure}\centering
\includegraphics[width=0.5\textwidth]{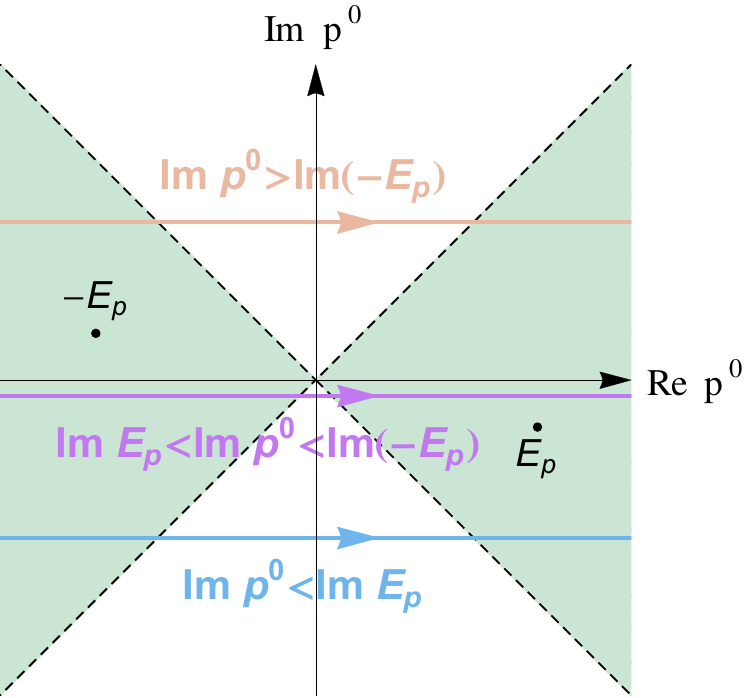}
\caption{
Shaded region represents convergent directions for $\ab{p^0}\to\infty$
for the integrand in $I\fn{\bs p}$.
The points $\pm\E_{\bs p}$ denote the poles of the propagator of $\Phi$.
The orange, purple, and blue lines represent the integral path 
for $\Im p^0>\Im\fn{-\E_{\bs p}}$, $\Im\E_{\bs p}<\Im p^0<\Im\fn{-\E_{\bs p}}$, and $\Im p^0<\Im\E_{\bs p}$, respectively.
\label{converging direction}}
\end{figure}

As shown in Fig.~\ref{converging direction}, the original $I\fn{\bs p}$ is a sum of $I_*\fn{\bs p}$ and the residue at the poles $p^0=\E_{\bs p}$ and $-\E_{\bs p}$ when $\Im p^0<\Im\E_{\bs p}$ and $\Im p^0>-\Im\E_{\bs p}$, respectively:
\al{
e^{-{\pn{\delta\mf T}^2\ov2\vs_+}}
I\fn{\bs p}
	&=
		e^{-{\pn{\delta\mf T}^2\ov2\vs_+}}I_*\fn{\bs p}
		+{e^{-{\vs_+\ov2}\sqbr{\omega_\vs\pn{\bs p}-\E_{\bs p}}^2+i\sqbr{\omega_\vs\pn{\bs p}-\E_{\bs p}}\delta\mf T}\ov2\E_{\bs p}}
			\theta\fn{\delta\mf T+\vs_+\Im\E_{\bs p}\ov\sqrt{2\stout}}\nn
	&\quad
		+{e^{-{\vs_+\ov2}\sqbr{\omega_\vs\pn{\bs p}+\E_{\bs p}}^2+i\sqbr{\omega_\vs\pn{\bs p}+\E_{\bs p}}\delta\mf T}\ov2\E_{\bs p}}
			\theta\fn{\vs_+\Im\E_{\bs p}-\delta\mf T\ov\sqrt{2\stout}},
			\label{Residue result}
}
where we have multiplied a positive constant $\vs_+\ov\sqrt{2\stout}$ in the Heaviside function to match Eq.~\eqref{emerging 1 to 2}.
The first and second terms in Eq.~\eqref{Residue result} correspond to the second and first terms in the braces in Eq.~\eqref{emerging 1 to 2}, respectively.
The last term in Eq.~\eqref{Residue result} corresponds to the ``backward propagation'' term dropped in Eq.~\eqref{emerging 1 to 2}.

We comment that the Wick rotation for the integration contour of $p^0$ from the real to imaginary axes can be justified only for the pole contributions, the second and third terms in Eq.~\eqref{Residue result}, whereas for the first term representing the wave-packet effect, the Wick rotation cannot be justified, as is obvious from the shaded region in Fig.~\ref{converging direction}. 
In other words, the Wick rotation becomes possible if we ignore the quadratic term with respect to $p^0$ in $f\fn{p^0;\bs p}$ by taking the limit $\vs_+\to\infty$ \emph{before} performing the $p^0$ integral, so that only the pole structure becomes relevant.

\subsubsection{Non-relativistic approximation}
So far in this section, we have left the integral over $\bs p$ untouched.
For comparison with Eq.~\eqref{limiting behavior of boundary and bulk} from Order~\ref{order 1}, we perform the $\bs p$-integral in Eq.~\eqref{S-matrix after 8 integrals}, that is, in Eq.~\eqref{wave-packet Feynman propagator equation} with Eq.~\eqref{Residue result}.
We will see that the results agree with each other.

In order to perform the integral around a saddle point $\bs p_*$ (in the limit $\sigma_+M^2\gg1$), we employ a non-relativistic approximation $\bs p\approx M\bs v$ and $E_{\bs p}\approx M+{M\bs v^2\ov2}$, which is valid when $\ab{\bs p_*}\ll M$:
\al{
\bs p
	&=	M\bs v+\Or{\ab{\bs v}^3},\\
E_{\bs p}
	&=	
		\sqrt{M^2+\bs p^2}
	=	
		M\pn{1+{\bs v^2\ov2}}+\Or{\ab{\bs v}^4},\\
\E_{\bs p}
	&=	\sqrt{\M^2+\bs p^2}
	=	\M\sqrt{1+{M^2\ov\M^2}\bs v^2}
	=	\M+{M^2\ov2\M}\bs v^2+\Or{\ab{\bs v}^4},\\
{\bs p\ov\E_{\bs p}}
	&=	{M\bs v\ov\M}+\Or{\ab{\bs v}^3},\\
\omega_\vs\fn{\bs p}
	&=	\omega_\vs\fn{0}
		+M\ol{\bs V}_\vs\cdot\bs v+\Or{\ab{\bs v}^3},
}
where $\bs v:={\bs p\ov E_{\bs p}}={\bs p\ov M}+\Or{\bs p^3}$ and $\M:=\sqrt{M^2-i\epsilon}$.
Then the exponent in the second term in Eq.~\eqref{I integral} becomes
\al{
&-{1\ov2}\pn{M\bs v-\bs p_*}_i
			\pn{\sigma_+\delta_{ij}+\vs\pn{\delta\ol{\bs V}}_i\pn{\delta\ol{\bs V}}_j}
			\pn{M\bs v-\bs p_*}_j
		-{\pn{\delta\mf T}^2\ov2\vs_+}\nn
	&
		-{\vs\ov2\pn{\sigma_++\vs\pn{\delta\ol{\bs V}}^2}}\pn{
			{\pn{\delta\ol{\bs V}}^2\pn{\ol{\bs\Xi}_\tx{out}-\ol{\bs\Xi}_\tx{in}}^2-\pn{\delta\ol{\bs V}\cdot\delta\ol{\bs\Xi}}^2\ov\sigma_+}
			+{\pn{\delta\ol{\bs\Xi}+\mf T_\vs\delta\ol{\bs V}}^2\ov\vs}
			}\nn
	&
		-{\vs_+\ov2}\pn{
			\omega_\vs\fn{0}
			-\M
			-i{\delta\mf T\ov\vs_+}}^2
		-\vs_+M\pn{
				\omega_\vs\fn{0}
				-\M
				-i{\delta\mf T\ov\vs_+}}
			\pn{
				\ol{\bs V}_\vs\cdot\bs v
			}
		+{\vs_+M\bs v^2\ov2}\pn{
				\omega_\vs\fn{0}
				-\M
				-i{\delta\mf T\ov\vs_+}}
				\nn
	&
		-{\vs_+M^2\ov2}\pn{\ol{\bs V}_\vs\cdot\bs v}^2
		-{\sigma\ov2}\pn{\bs P_\tx{out}-\bs P_\tx{in}}^2
		-{\vs\sigma_+\ov2\pn{\sigma_++\vs\pn{\delta\ol{\bs V}}^2}}
			\pn{E_\tx{out}-E_\tx{in}-\ol{\bs V}_\sigma\cdot\pn{\bs P_\tx{out}-\bs P_\tx{in}}}^2\nn
	&
		+\Or{\ab{\bs v}^3},
		\label{exponent intermediate}
}
where
the following are the weighted averages
\al{
\ol{\bs V}_\sigma
	&:=	\sigma\pn{
			{\ol{\bs V}_\tx{in}\ov\sigma_\tx{in}}
			+{\ol{\bs V}_\tx{out}\ov\sigma_\tx{out}}
			},&
\ol{\bs V}_\vs
	&:=	
			{\stin\ol{\bs V}_\tx{in}+\stout\ol{\bs V}_\tx{out}\ov\stin+\stout},&
\mf T_\vs
	&:=	\vs\pn{{\mf T_\tx{in-int}\ov\stin}+{\mf T_\tx{out-int}\ov\stout}},
}
and
\al{
\delta\ol{\bs V}
	&:=	\ol{\bs V}_\tx{out}-\ol{\bs V}_\tx{in},\\
\bs p_*
	&=	\paren{
			\bs P_\sigma
			-{\varsigma\,\delta\omega\fn{0}\,\delta\ol{\bs V}-i\paren{\delta\ol{\bs\Xi}+\mf T_\varsigma\,\delta\ol{\bs V}}\ov\sigma_+}
			}\nn
	&\quad
		-{\varsigma\paren{\delta\ol{\bs V}}^2\ov\sigma_++\varsigma\paren{\delta\ol{\bs V}}^2}\paren{
			\bs P_\sigma
			-{\varsigma\,\delta\omega\fn{0}\,\delta\ol{\bs V}-i\paren{\delta\ol{\bs\Xi}+\mf T_\varsigma\,\delta\ol{\bs V}}\ov\sigma_+}
			}_\parallel,
}
in which
\al{
\delta\omega\fn{0}
	&:=	\omega_\tx{out}\fn{0}-\omega_\tx{in}\fn{0},\\
\bs Q_\parallel
	&:=	{\paren{\delta\ol{\bs V}\cdot\bs Q}\ov\paren{\delta\ol{\bs V}}^2}\delta\ol{\bs V}.
}
Note that
\al{
\bs p_*\cdot\delta\ol{\bs V}
	&=	{\sigma_+\ov\sigma_++\vs\pn{\delta\ol{\bs V}}^2}\paren{
			\bs P_\sigma
			-{\varsigma\,\delta\omega\fn{0}\,\delta\ol{\bs V}-i\paren{\delta\ol{\bs\Xi}+\mf T_\varsigma\,\delta\ol{\bs V}}\ov\sigma_+}
			}\cdot\delta\ol{\bs V}.
}

Now we can perform the Gaussian integral over $\bs v$. The saddle point is located at
\al{
\bs v_\star
	&=	{1\ov
			-\gamma
				}\sqbr{
		\bs\delta
		-{
			\pn{\bs a\cdot\bs \beta}\pn{\bs b\cdot\bs \delta}
			-\pn{\bs a\cdot\bs\delta}\pn{\gamma+\bs b\cdot\bs \beta}
			\ov
				\pn{\bs a\cdot\bs\beta}\pn{\bs b\cdot\bs\alpha}
				-\pn{\gamma+\bs a\cdot\bs\alpha}\pn{\gamma+\bs b\cdot\bs\beta}
			}\bs\alpha
		-{
			\pn{\bs a\cdot\bs\delta}\pn{\bs b\cdot\bs\alpha}
			-\pn{\bs b\cdot\bs\delta}\pn{\gamma+\bs a\cdot\bs\alpha}
				\ov
					\pn{\bs a\cdot\bs\beta}\pn{\bs b\cdot\bs\alpha}
					-\pn{\gamma+\bs a\cdot\bs\alpha}\pn{\gamma+\bs b\cdot\bs\beta}
			}\bs\beta
		},
}
where
\al{
\bs a
	&=	\delta\ol{\bs V},&
\bs b
	&=	\ol{\bs V}_\vs,&
\bs\alpha
	&=	-\vs M^2\delta\ol{\bs V},&
\bs\beta
	&=	-\vs_+M^2\ol{\bs V}_\vs,
}
and
\al{
\gamma
	&=	-\sigma_+M^2+{\vs_+M^2\ov\M}\pn{\omega_\vs\fn{0}-\M-i{\delta\mf T\ov\vs_+}},\\
\bs\delta
	&=	\vs M\pn{\bs p_*\cdot\delta\ol{\bs V}}\delta\ol{\bs V}
		-\vs_+M\ol{\bs V}_\vs\pn{\omega_\vs\fn{0}-\M-i{\delta\mf T\ov\vs_+}}.
}
The three eigenvalues of the Hessian matrix are
\al{
-\sqbr{
			M^2\sigma_+
			+{\vs_+M^2\ov\M}\pn{
				\omega_\vs\fn{0}
				-\M
				-i{\delta\mf T\ov\vs_+}}
			}
}
and
\al{
&-\sqbr{
			M^2\sigma_+
			+{\vs_+M^2\ov\M}\pn{
				\omega_\vs\fn{0}
				-\M
				-i{\delta\mf T\ov\vs_+}}
			}
-{\vs M^2\pn{\delta\ol{\bs V}}^2+\vs_+M^2\ol{\bs V}_\vs^2\ov2}\nn
&\quad
\pm{M^2\sqrt{
		\pn{\vs \pn{\delta\ol{\bs V}}^2-\vs_+\ol{\bs V}_\vs^2}^2
		+4\vs\vs_+\pn{\delta\ol{\bs V}\cdot\ol{\bs V}_\vs}^2
		}\ov2}.
}

For the simple configuration~\eqref{simple configuration}, we get
\al{
\bs v_\star
	&\to
		0,\\
\bs p_*
	&\to	\bs P_\sigma+i{\delta\ol{\bs\Xi}\ov\sigma_+},\\
\tx{Eq.~\eqref{exponent intermediate}}
	&\to
		-{\sigma_+\ov2}\pn{\bs P_\sigma+i{\delta\ol{\bs\Xi}\ov\sigma_+}}^2
		-{\pn{\delta\mf T}^2\ov2\vs_+}
		-{\pn{\ol{\bs\Xi}_\tx{out}-\ol{\bs\Xi}_\tx{in}}^2\ov2\sigma_+}\nn
	&\quad
		-{\vs_+\ov2}\pn{
			\omega_\vs\fn{0}
			-\M
			-i{\delta\mf T\ov\vs_+}}^2,\\
(\tx{All the three eigenvalues})
	&\to
		-\sqbr{
			M^2\sigma_+
			+{\vs_+M^2\ov\M}\pn{
				E_\vs
				-\M
				-i{\delta\mf T\ov\vs_+}}
			},
}
which results in
\al{
\mc S
	&\to
			\pn{2\pi}^4N
				\sqbr{\pn{\sigma\ov2\pi}^{3/2}e^{-{\sigma\ov2}\pn{\bs P_\tx{out}-\bs P_\tx{in}}^2}}
				\sqbr{
					\sqrt{\vs\ov2\pi}e^{-{\vs\ov2}\pn{E_\tx{out}-E_\tx{in}}^2}
					}
				e^{i\pn{E_\vs-\M}\delta\mf T}\nn
	&\quad\times{\sqrt{2\pi\vs_+}\ov2\M}\Bigg\{
			\theta\fn{\delta\mf T+\vs_+\Im\M}
			{
				e^{
				-{\sigma_+\ov2}\bs P_\sigma^2
				-{\vs_+\ov2}\pn{\omega_\vs\fn{0}-\M}^2
				-i\bs P_\sigma\cdot\delta\ol{\bs\Xi}
				}
				\ov\pn{
			1
			+{\vs_+\ov\sigma_+}{
				\omega_\vs\fn{0}-\M
				-i{\delta\mf T\ov\vs_+}\ov\M}
			}^{3/2}}
		-{1\ov\sqrt{2\pi\vs_+}}{-ie^{
			-{\paren{\delta\mf T}^2\ov2\varsigma_+}
			-{\paren{\delta\ol{\bs\Xi}}^2\ov2\sigma_+}
			-i\pn{E_\vs-\M}\delta\mf T
			}
			\ov
			\omega_\vs\fn{0}-\M-i{\delta\mf T\ov\vs_+}
			}
		\Bigg\}.
		\label{NR result}
}
The full expression corresponding to Eq.~\eqref{NR result} can be easily obtained without taking the simple configuration~\eqref{simple configuration}, but it is lengthy and will be omitted.

We see that Eq.~\eqref{NR result} matches Eq.~\eqref{limiting behavior of boundary and bulk} with the relation $\M\approx\E_\sigma$ and $\omega_\vs\fn{0}= E_\vs$ ($\to E_\tx{out}$).
Note that the consistency condition $\ab{\bs p_*}\ll M$ for the non-relativistic approximation requires ${\delta\ol{\bs\Xi}}^2/\sigma_+^2\ll M^2$ along with $\bs P_\sigma^2\ll M^2$ and hence the factors $e^{-{(\delta\ol{\bs\Xi})^2/2\sigma_+}}$, $e^{-{\sigma_+\ov2}\bs P_\sigma^2}$, and $e^{-i\bs P_\sigma\cdot\delta\ol{\bs\Xi}}$ are irrelevant.

\subsection{(b) Lefschetz thimble decomposition}\label{approach b section}

\begin{figure}\centering
\includegraphics[width=0.5\textwidth,clip]{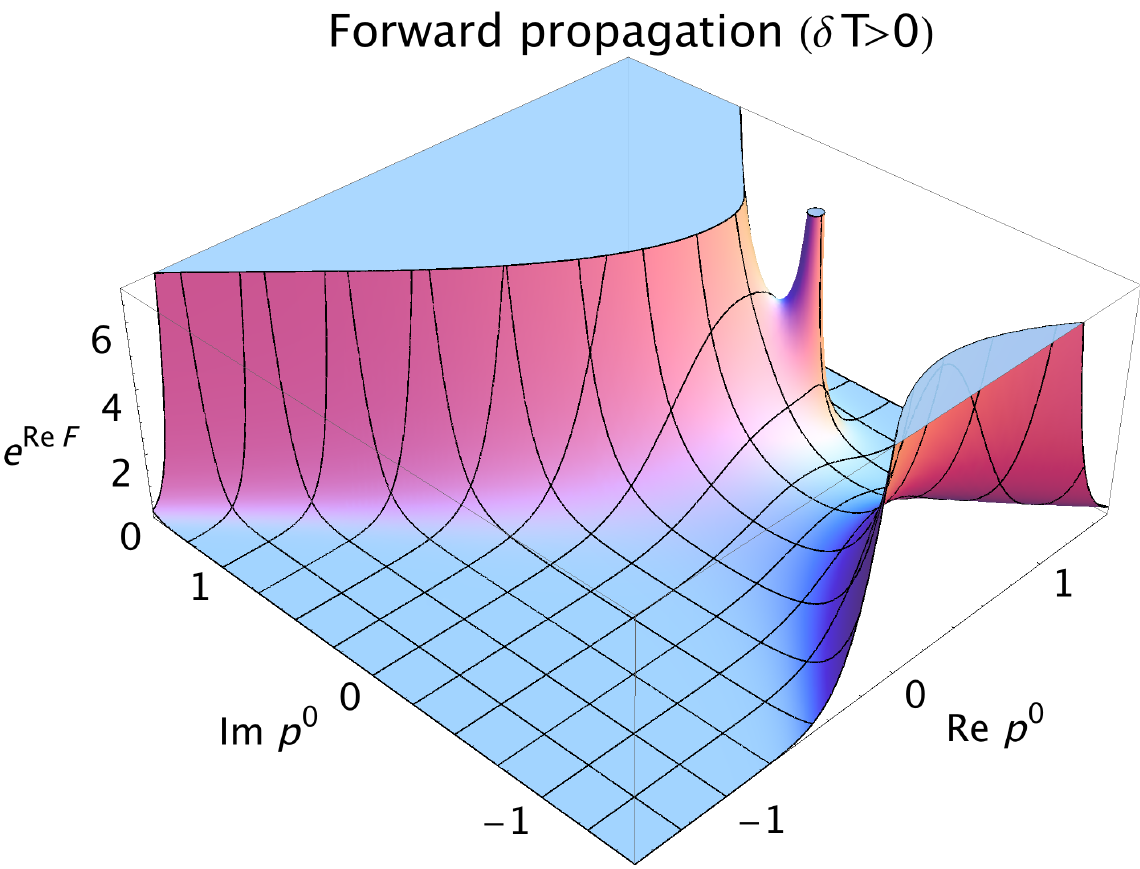}
\caption{
We plot $e^{\Re\mc F}$ with sample parameters $\vs_+=10$, $\omega_\vs\fn{\bs p_*}$ (=$\Re p^0_*$) $=0.5$, and $\epsilon=0.3$ in the $E_{\bs p}=1$ units. This corresponds to forward propagation of $\Phi$ with ${\delta\mf T\ov\vs_+}$ ($=-\Im p^0_*$) $=0.5$. 
\label{3D figure}
}
\end{figure}


Here, we evaluate the $p^0$-integral in Eq.~\eqref{S-matrix after 8 integrals} by the Lefschetz thimble decomposition to confirm Eq.~\eqref{Residue result}.
That is, we show that the result~\eqref{Residue result} can be understood as a Stokes phenomenon.

We first rewrite the integral~\eqref{I appearing again} into an exponential form:
\al{
I\fn{\bs p}
	&=	\int_{-\infty}^\infty{\df p^0\ov2\pi i}e^{\mc F\pn{p^0;\bs p}},
}
where
\al{
\mc F\fn{p^0;\bs p}
	&=	-{\vs_+\ov2}\pn{p^0-p^0_*\fn{\bs p}}^2-\ln\fn{-\pn{p^0}^2+\E_{\bs p}^2}.
		\label{curly F given}
}
Recall that $p^0_*\fn{\bs p}$ is given in Eq.~\eqref{p0star given}.
Hereafter, we suppress the $\bs p$-dependence and write $\mc F\fn{p^0}$, etc.\ for simplicity.
As an illustration, we plot the absolute value of the integrand $e^{\Re\mc F}$ on the complex $p^0$ plane in Fig.~\ref{3D figure}, with sample parameters $\vs_+=10$, $\omega_\vs\fn{\bs p_*}$ (=$\Re p^0_*$) $=0.5$, and $\epsilon=0.3$ in the $E_{\bs p}=1$ units. 
The second logarithmic term in Eq.~\eqref{curly F given} modifies the exponent but the qualitative pole structure remains the same.

By solving ${\p\mc F\ov\p p^0}=0$, we obtain the three saddle points $p^0_{(*)}$ and $p^0_{(\pm)}$ that are shifted from $p^0_*$ and $\pm\E_{\bs p}$ (see Fig.~\ref{converging direction}) due to the extra logarithmic term in $\mc F\fn{p^0}$. 
Concretely, the saddle-point equation reads
\al{
{\p\mc F\ov\p p^0}
	&=	-\vs_+\pn{p^0-p^0_*\fn{\bs p}}-{-2p^0\ov-\pn{p^0}^2+\E_{\bs p}^2}
	=	0.
}
This is cubic in $p^0$, and we obtain exact solutions for the saddle points $p^0_{(i)}$.
Here we show the large-$\vs_+$ results as an illustration, though we use the exact ones in the numerical plots below:
\al{
p^0_{(*)}
	&\simeq
		p^0_*+{1\ov\vs_+}{
			{2p^0_*\ov-\pn{p^0_*}^2+\E_{\bs p}^2}
			},&
p^0_{(\pm)}
	&\simeq
		\pm\E_{\bs p}
		+{1\ov\vs_+}{
			{1\ov p^0_*\mp\E_{\bs p}}
			},
			\label{approximate Lefschetz saddle point}
}
with
\al{
\mc F_{(*)}
	&\simeq
		\ln{1\ov-\pn{p^0_*}^2+\E_{\bs p}^2}
		+{1\ov\vs_+}{2\pn{p^0_*}^2\ov\pn{\pn{p^0_*}^2-\E_{\bs p}^2}^2}
		,\\
\mc F_{(\pm)}
	&\simeq
		-{\vs_+\ov2}\pn{p^0_*\mp\E_{\bs p}}^2
		+\ln\fn{{\vs_+\ov2}\pn{1\mp{p^0_*\ov\E_{\bs p}}}}
		+1,
}
and
\al{
{\p^2\mc F_{(*)}\ov{\p p^0}^2}
	&\simeq
		-\vs_+
		+{1\ov\pn{p^0_*+\E_{\bs p}}^2}
		+{1\ov\pn{p^0_*-\E_{\bs p}}^2}
		,\\
{\p^2\mc F_{(\pm)}\ov{\p p^0}^2}
	&\simeq
		\vs_+^2\pn{p^0_*\mp\E_{\bs p}}^2
		-\pn{2\pm{p^0_*\ov\E_{\bs p}}}\vs_+
		,
}
where we have shown up to the sub-leading terms (as well as the order-$\ln\vs_+$ term) for large $\vs_+$.

For each saddle point $(i)$ with $i=*$ and $\pm$, the steepest decent and ascent paths are obtained from the condition $\Im\Fn{\mc F\fn{p^0}-\mc F\Fn{p^0_{(i)}}}=0$.
The steepest descent path $\mc J_{(i)}$ (ascent path $\mc K_{(i)}$) from the saddle point $(i)$ is called the Lefschetz thimble or the Stokes line (the anti-thimble or the anti-Stokes line).
The integral path in the pole approach (a), which is one of the colored horizontal lines in Fig.~\ref{converging direction}, corresponds to $\mc J_{(*)}$ in the current approach (b).
On the other hand, the poles $\pm\E_{\bs p}$ in Fig.~\ref{converging direction} correspond to the contour-integral along $\mc J_{(\pm)}$, respectively.

\begin{figure}\centering\hfill
\includegraphics[width=0.3\textwidth]{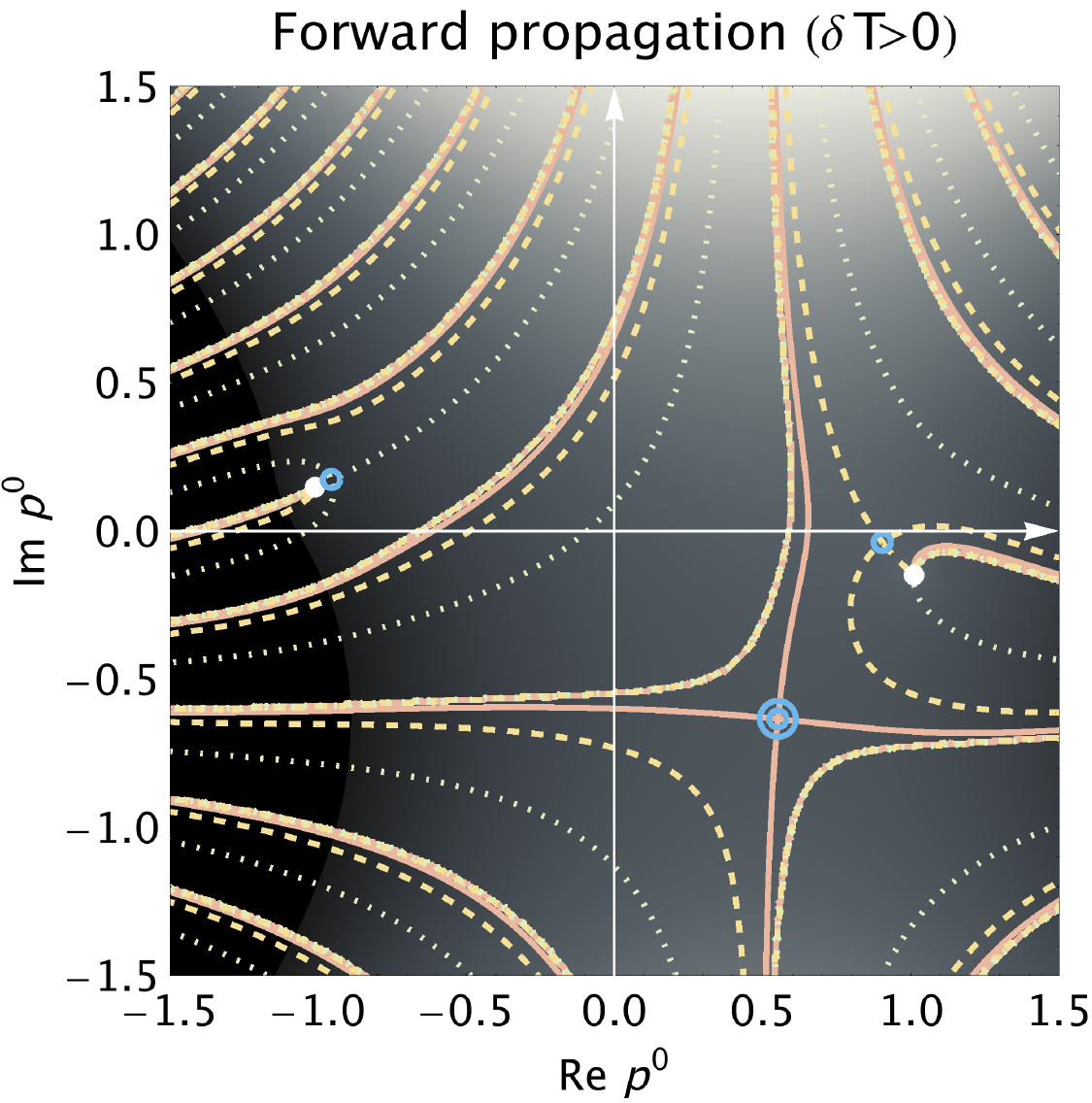}\hfill
\includegraphics[width=0.3\textwidth]{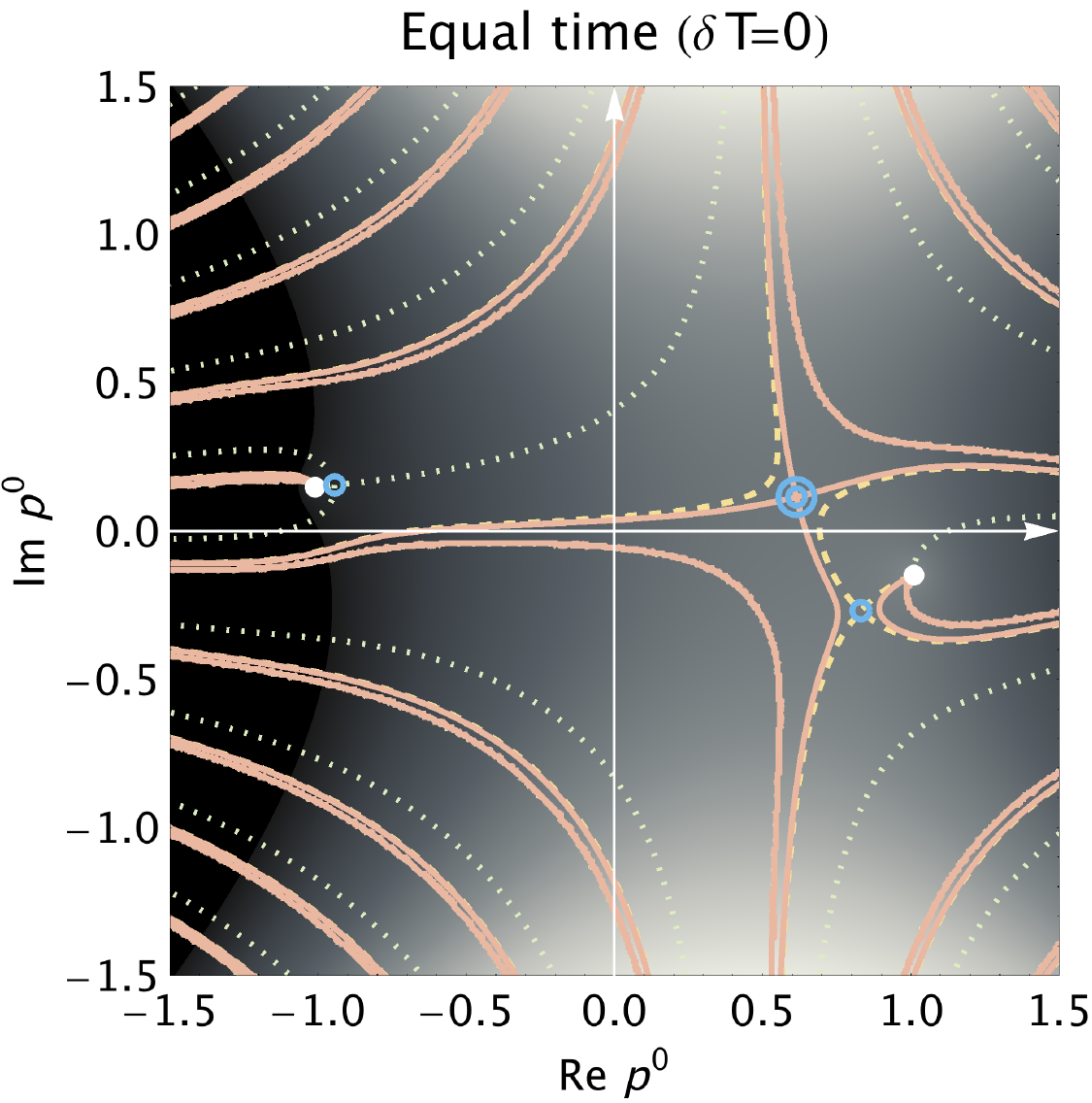}\hfill
\includegraphics[width=0.3\textwidth]{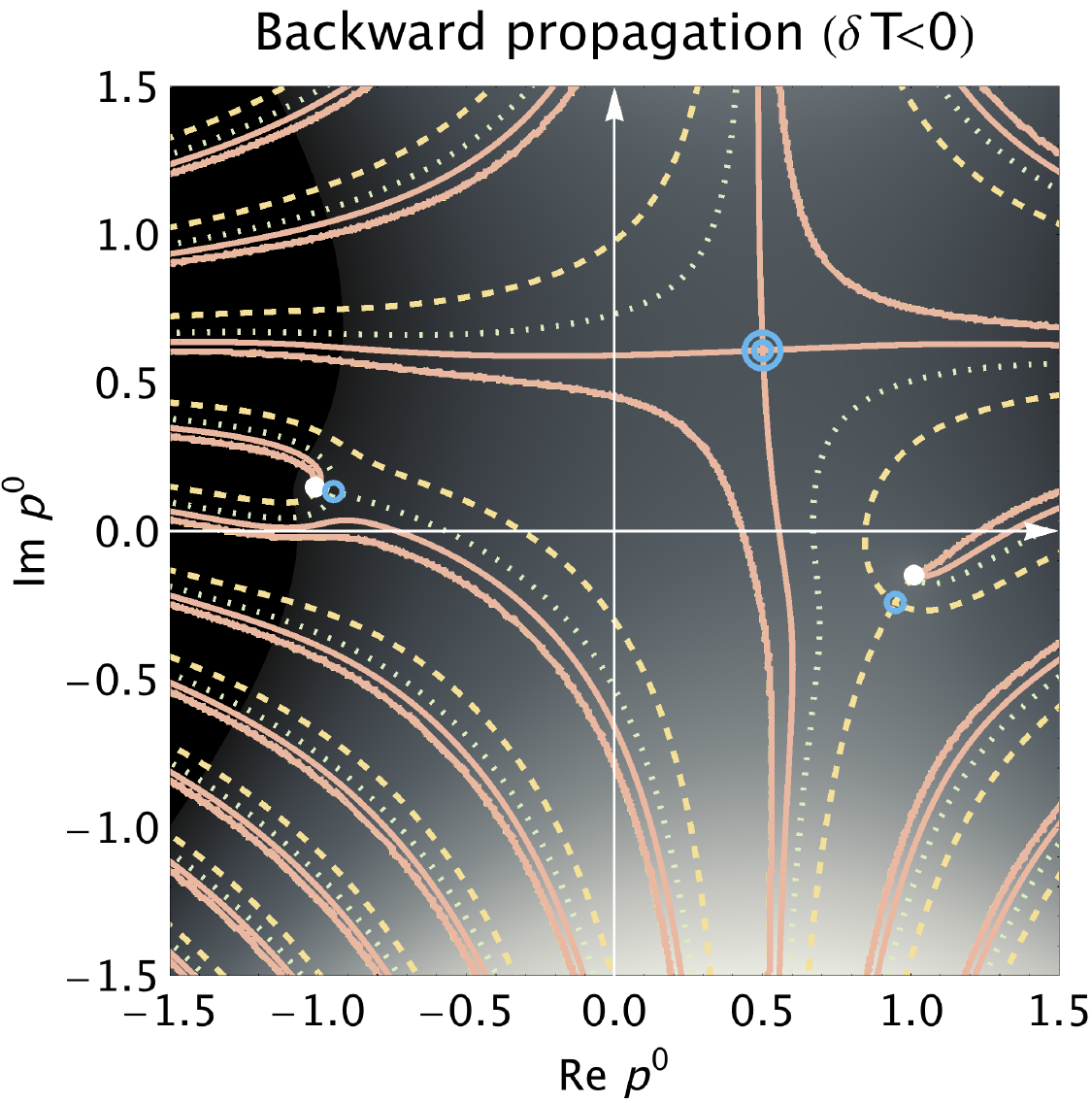}\hfill
\includegraphics[width=0.053\textwidth]{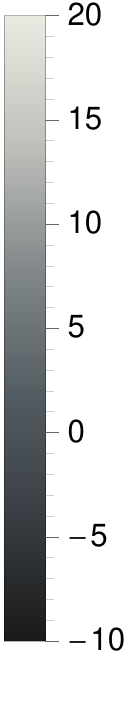}\hfill\mbox{}\\
\caption{We show contours for $\Im\mc F=0$ (mod $2\pi$), on a density plot of $\Re\mc F$ in $[-10,20]$. We have chosen sample parameters $\vs_+=10$, $\omega_\vs\fn{\bs p_*}$ (=$\Re p^0_*$) $=0.5$, and $\epsilon=0.3$ in the $E_{\bs p}=1$ units. The first panel corresponds to forward propagation of $\Phi$ with ${\delta\mf T\ov\vs_+}$ ($=-\Im p^0_*$) $=0.5$, while the second and third to equal-time and backward with ${\delta\mf T\ov\vs_+}=0$ and $-0.5$, respectively.
The red-solid, yellow-dashed, and green-dotted contours are $\Im\mc F=\Im\mc F_{(*)}$, $\Im\mc F_{(+)}$, and $\Im\mc F_{(-)}$ respectively. 
The white dots denote the poles at $p^0=\pm\E_{\bs p}$ and the blue single circles nearby them are correspondingly the saddle points $p^0_{(\pm)}$.
The blue double circle denotes the saddle point $p^0_{(*)}$.
\label{stokes figure}
}
\end{figure}

As an illustration, we plot in Fig.~\ref{stokes figure} contours for $\Im\mc F=0$ (mod $2\pi$), superimposed on the density plot for $\Re\mc F$.
The blue (double) circles denote saddle points $p^0_{(i)}$, and the white dots denote the poles $\pm\E_{\bs p}$.
The horizontal and vertical contours (both red-solid) through $p^0_{(*)}$ are the thimble $\mc J_{(*)}$ and the anti-thimble $\mc K_{(*)}$, respectively.
Through $p^0_{(+)}$, the yellow-dashed contour that connects $\E_{\bs p}$ and a $\ab{\Im p^0}\to\infty$ region is the anti-thimble $\mc K_{(+)}$, while its perpendicular contour is the thimble $\mc J_{(+)}$.
Through $p^0_{(-)}$, the green-dotted contour connecting $-\E_{\bs p}$ and a $\ab{\Im p^0}\to\infty$ region is $\mc K_{(-)}$, and its perpendicular, $\mc J_{(-)}$.
All the other contours that are disconnected from any of $p^0_{(i)}$ are irrelevant to the integral.

\begin{figure}\centering
\includegraphics[width=0.3\textwidth,clip]{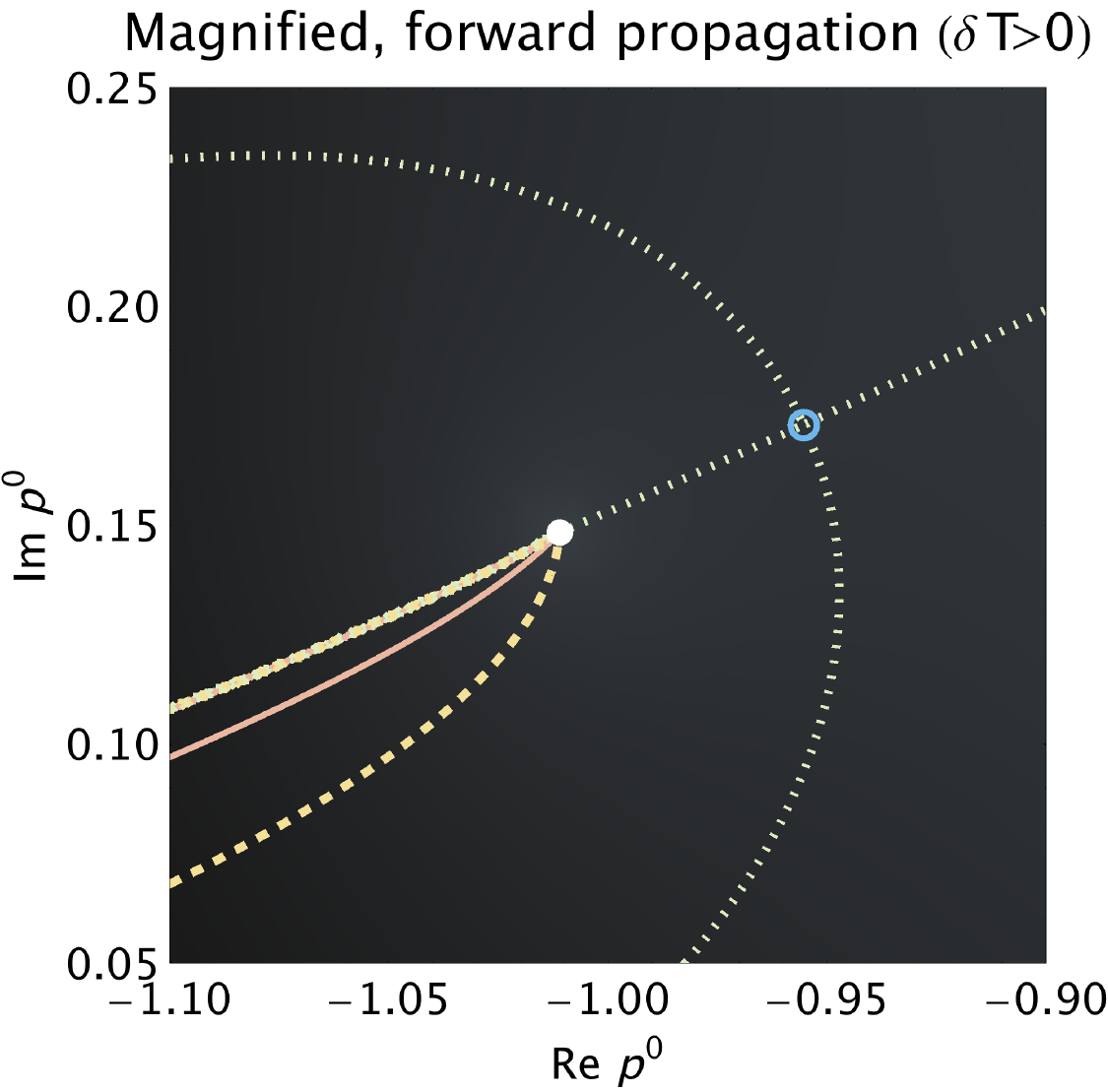}
\caption{
Magnified plot for $\Im\mc F$ (mod $2\pi$) of the second panel in Fig.~\ref{stokes figure} near the saddle point $p^0_{(-)}$ (blue circle) and the pole $-\E_{\bs p}$ (white dot). We see that three contours and an artifact line terminate in $-\E_{\bs p}$: The three contours are $\Im\mc F=\Im\mc F_{(*)}$ (red-solid), $\Im\mc F=\Im\mc F_{(+)}$ (yellow-dashed), and the anti-thimble $\mc K_{(-)}$ (green-dotted) from $p^0_{(-)}$. The extra artifact line, seemingly consisting of three degenerate contours, is from the branch cut of the logarithmic function in the numerical computation, along which the contributions from all the other Riemann surfaces appear. This way, in each pole in Fig.~\ref{stokes figure}, there always terminate three contours and an artifact line.
\label{magnified figure}
}
\end{figure}

The figure is drawn on a single Riemann surface for the logarithmic function in $\mc F\fn{p^0}$:
In the case $\Im p^0_{(*)}\lesssim\Im\E_{\bs p}$ in the first panel in Fig.~\ref{stokes figure}, we may put a branch cut from $-\E_{\bs p}$ to $-\infty+\Im\fn{-\E_{\bs p}}$ and another from $\E_{\bs p}$ along a, say, red-solid or green-dotted contour. This way, we may continuously deform $\mathbb R$ into the thimbles $\mc J_{(*)}$ and $\mc J_{(+)}$ without crossing the cuts. We may do similarly for other cases.
In Fig.~\ref{magnified figure}, we explain artifact lines appearing in Fig.~\ref{stokes figure} due to the branch cuts in the numerical computation.

We take the direction of $\mc J_{(\pm)}$ such that it can be deformed to $\mathbb R$ without crossing the poles, namely, $\mc J_{(+)}$ and $\mc J_{(-)}$ circulate $\E_{\bs p}$ and $-\E_{\bs p}$ clockwise and counterclockwise, respectively.
Along $\mc J_{(i)}$, we may evaluate the approximate Gaussian integral
\al{
I_{(i)}\fn{\bs p}
	&=	\int_{\mc J_{(i)}}{\df p^0\ov2\pi i}\,e^{\mc F\pn{p^0;\bs p}}
}
without the oscillation of integrand:
\al{
I_{(*)}\fn{\bs p}
	&\simeq
		{1\ov\sqrt{2\pi\vs_+}}{-i\ov-\pn{p^0_*\fn{\bs p}}^2+\E_{\bs p}^2},\\
I_{(\pm)}\fn{\bs p}
	&\simeq
		\mp{-i\ov\sqrt{2\pi}}{1\ov\sqrt{-\vs_+^2\pn{p^0_*\fn{\bs p}\mp\E_{\bs p}}^2}}e^{-{\vs_+\ov2}\pn{p^0_*\fn{\bs p}\mp\E_{\bs p}}^2+1}{\vs_+\ov2}\pn{1\mp{p^0_*\fn{\bs p}\ov\E_{\bs p}}}
	=	{e\ov\sqrt{2\pi}}{e^{-{\vs_+\ov2}\pn{p^0_*\fn{\bs p}\mp\E_{\bs p}}^2}\ov2\E_{\bs p}},
	\label{I plus minus}
}
where we have taken $\sqrt{re^{i\theta}}=\sqrt re^{i\theta/2}$ for $-\pi<\theta<\pi$, namely, $\sqrt{\pn{p^0_*\fn{\bs p}\mp\E_{\bs p}}^2}=\E_{\bs p}\mp p^0_*\fn{\bs p}$.

The integral $I$ can be decomposed into those on the Lefschetz thimbles:
\al{
I	=	\sum_{i=*,\pm}\Braket{\mc K_{(i)},\mathbb R}I_{(i)},
	\label{thimble decomposition}
}
where $\Braket{\mc K_{(i)},\mathbb R}$ is the intersection number between the anti-thimble $\mc K_{(i)}$ and the original integration path $\mathbb R$; see e.g.\ Ref.~\cite{Tanizaki:2015gpl} for a review.
This expression allows an interpretation of the appearance of the Heaviside step functions in Eq.~\eqref{Residue result} as a Stokes phenomenon, which we discuss from now on.

There are three cases depending on the relative position of $\mc J_{(*)}$ and $\pm\E_{\bs p}$, represented in Fig.~\ref{stokes figure}:
\begin{enumerate}[(I)]
\item When $\Im p^0_{(*)}\lesssim\Im\E_{\bs p}$ as in the first panel, the anti-thimble $\mc K_{(-)}$ terminates in $-\E_{\bs p}$, and hence do not intersect with the real axis: $\Braket{\mc K_{(-)},\mathbb R}=0$;
see Fig.~\ref{magnified figure} for magnification and more explanation.
\item When $\Im\E_{\bs p}\lesssim\Im p^0_{(*)}\lesssim\Im\fn{-\E_{\bs p}}$ as in the second panel, both anti-thimbles terminate in $\pm\E_{\bs p}$, and hence $\Braket{\mc K_{(\pm)},\mathbb R}=0$.
\item When $\Im p^0_{(*)}\gtrsim\Im\fn{-\E_{\bs p}}$ as in the third panel, $\mc K_{(+)}$ terminates in $\E_{\bs p}$, and hence $\Braket{\mc K_{(+)},\mathbb R}=0$.
This is how the discrete change in the amplitude~(4) in the main text is understood as a Stokes phenomenon.
Cases (I), (II), and (III) correspond to $\Im p^0<\Im\E_{\bs p}$, $\Im\E_{\bs p}<\Im p^0<\Im\fn{-\E_{\bs p}}$, and $\Im p^0>\Im\fn{-\E_{\bs p}}$ in Approach~(a), respectively.
\end{enumerate}


We see that the results of Approaches~(a) and (b) in Order~\ref{order 2}, namely Eqs.~\eqref{Residue result} and \eqref{thimble decomposition}, respectively obtained in this Sec.~\ref{Order 2 section}, are identical up to the extra factor ${e\ov\sqrt{2\pi}}\simeq1.08$ that appears in Eq.~\eqref{I plus minus}.\footnote{
It would be interesting to pursue the difference of the factor $e/\sqrt{2\pi}\simeq1.08$ between Approaches~(a) and (b).
In the former, the saddle-point integral is done on the straight line, while in the latter, the integral on the thimble is approximated by the Gaussian integral on the straight line that is tangent to the thimble, which is highly bent in the limit $\vs_+\to\infty$.
This might be a cause of the deviation.
}

\subsection{Discussion on wave-packet Feynman propagator}\label{Wave-packet Feynman propagator section}
We comment on the pole structure of the wave-packet Feynman propagator~\eqref{wave-packet Feynman propagator}.
From the dependence on $\delta\mf T$ for a fixed $\vs_+$, as well as on the energy difference such as $E_{\bs p}-\omega_\vs\fn{\bs p}$, we identify the first term in Eq.~\eqref{Residue result} as the time-boundary effect, while the second and third ones as the bulk effect that has a counterpart in the plane-wave propagator:~\footnote{\label{opposite footnote}
This is opposite to the interpretation in Ref.~\cite{Ishikawa:2020hph}.
}
\begin{itemize}
\item In the first term,
(i)~the real and imaginary parts of the energy pole are shifted as $E_{\bs p}\to E_{\bs p}-\omega_\vs\fn{\bs p}$ and $-{\epsilon\ov2E_{\bs p}}\to-{\epsilon\ov2E_{\bs p}}+{\delta\mf T\ov\vs_+}$, respectively;
(ii)~the energy non-conservation is suppressed only by a power law; and
(iii)~there is exponential suppression for large $\delta\mf T$.
These are all the characteristics of the time boundary effect~\cite{Ishikawa:2018koj}.
\item In the second and third terms,
(i)~the pole structure ${1\ov\E_{\bs p}}$ is the same as the ordinary plane-wave propagator;
(ii)~the energy is exponentially localized to the positive and negative poles $\omega_\vs\fn{\bs p}\simeq\E_{\bs p}$ and $\omega_\vs\fn{\bs p}\simeq-\E_{\bs p}$, respectively; and
(iii)~when we first take the imaginary part infinitesimal as in the Feynman propagator, $\Im\E_{\bs p}\to0$, there is no exponential suppression for large $\delta\mf T$.
These are all the characteristics of the bulk effect~\cite{Ishikawa:2018koj}.
\end{itemize}


Historically in the original paper~\cite{Feynman:1949hz}, Feynman first considered a transition amplitude between a wave packet that is time-translated from $f\fn{\bs x_1}$ at $t=0$ to $T$ ($>0$) and a wave packet $g\fn{\bs x_2}$ at $t=T$ 
under an external potential $A$. Therefore the time integral over the interaction point in the resultant amplitudes~(22) and (23) in Ref.~\cite{Feynman:1949hz} must be between 0 and $T$~\footnote{\label{dressed}
This has to be the case unless we change the physical state to the ``dressed'' one in the sense of Ref.~\cite{Ishikawa:2020hph}. This change was made implicitly when Feynman gave the plane-wave amplitude in the momentum basis in the subsequent part in Ref.~\cite{Feynman:1949hz}.
}.
Then he implicitly changed the time-boundary conditions to extract a manifestly Lorentz-covariant amplitude, corresponding to a transition from a plane wave at time $-\infty$ to a one at $\infty$.
In the sequel paper~\cite{Feynman:1949zx}, he generalized this to many-body amplitudes {\it ``imagining that we can neglect the effect of interactions''} near the time boundaries, hoping that {\it ``we do not lose much in a general theoretical sense by this approximation''}; see also Ref.~\cite{Lippmann:1950zz}.
In the present paper, we calculate the amplitude including the time-boundary contribution that Feynman neglected.
It is important that the wave-packet form~\eqref{Residue result} is a finite complete expression without any singularity and \emph{contains} the plane-wave amplitude. 
We stress that the Gaussian basis forms a complete set that spans each free one-particle Hilbert space (see e.g.\ Refs.~\cite{Ishikawa:2005zc,Oda:2021tiv}), and hence the time-boundary effect must not be omitted.\footnote{
We comment that the standard textbook by Goldberger and Watson~\cite{GoldbergerWatson} also takes the same approach as that of Feynman, dropping the contributions from the asymptotic time region.
}


\section{Summary}\label{summary section}
We have elucidated the nature of the wave-packet scattering amplitude $\phi\phi\to\Phi\to\phi\phi$.
The wave-packet formalism provides a finite amplitude-squared that has no ambiguity of the delta-function-squared of the energy-momentum, unlike the plane-wave amplitude-squared.
The obtained amplitude provides a complete formula for the resonant production of $\Phi$ and its subsequent decay.

First, in Order~\ref{order 1}, we have explicitly computed the wave-packet scattering amplitude relying on the saddle-point approximation in Sec.~\ref{emergence of time boundary section}.
We have proven that the in-time-boundary term for the $\Phi\to\phi\phi$ decay amplitude \emph{emerges} from this scattering amplitude, even if we neglect both the in and out time boundaries for the $\phi\phi\to\phi\phi$ scattering process. 
For the first time, we have \emph{derived} the time boundary term shown in Eq.~\eqref{emerging 1 to 2} without introducing any time boundary a priori.
We stress that our formula is applicable to the general situation, including the intermediate state $\Phi$ being long-lived.

Second, in Order~\ref{order 2}, we have confirmed the result in the different integration order in Sec.~\ref{Order 2 section}.
The confirmation is done in the two different approaches~(a) saddle-point and poles and (b) Lefschetz thimble decomposition in Secs.~\ref{approach a section} and \ref{approach b section}, respectively. In Sec.~\ref{Wave-packet Feynman propagator section}, we discussed the relation to the ordinary plane-wave Feynman propagator in detail.
The wave-packet amplitude contains more information than the plane-wave counterpart and exhibits the time-boundary effect, which is hardly tractable in the plane-wave formulation.

The time-boundary effect directly influences the transition probability, a critical element of quantum physics, and hence has a great impact on vast sectors of scientific study.

\subsection*{Acknowledgement}
We thank Osamu Jinnouchi and Ryosuke Sato for the useful discussion and Juntaro Wada for reading the manuscript of the first draft.
This work is in part supported by JSPS Kakenhi Grant Nos.~21H01107 (K.I., K.N., and K.O.) and 19H01899 (K.O.).

\appendix
\section*{Appendix}

\section{Notations}\label{notation section}
We list the symbols defined in the main text:
\al{
\bs V_a
	&=	{\bs P_a\ov E_a}
	=	{\bs P_a\ov\sqrt{m^2+\bs P_a^2}},
}
\al{
E_{\bs p}
	&=	\sqrt{M^2+\bs p^2},&
\E_{\bs p}
	&=	\sqrt{E_{\bs p}^2-i\epsilon}
	=	\sqrt{M^2+\bs p^2-i\epsilon},&
\M
	&=	\sqrt{M^2-i\epsilon},
}
\al{
E_\tx{in}
	&=	E_1+E_2,&
E_\tx{out}
	&=	E_3+E_4,\\
\bs P_\tx{in}
	&=	\bs P_1+\bs P_2,&
\bs P_\tx{out}
	&=	\bs P_3+\bs P_4,\\
\sigma_\tx{in}
	&=	{1\ov{1\ov\sigma_1}+{1\ov\sigma_2}}
	=	{\sigma_1\sigma_2\ov\sigma_1+\sigma_2},&
\sigma_\tx{out}
	&=	{1\ov{1\ov\sigma_3}+{1\ov\sigma_4}}
	=	{\sigma_3\sigma_4\ov\sigma_3+\sigma_4},\\
\ol{\bs V}_\tx{in}
	&=	\sigma_\tx{in}\pn{{\bs V_1\ov\sigma_1}+{\bs V_2\ov\sigma_2}},&
\ol{\bs V}_\tx{out}
	&=	\sigma_\tx{out}\pn{{\bs V_3\ov\sigma_3}+{\bs V_4\ov\sigma_4}},\\
\omega_\tx{in}\fn{\bs p}
	&=	E_\tx{in}+\ol{\bs V}_\tx{in}\cdot\pn{\bs p-\bs P_\tx{in}},&
\omega_\tx{out}\fn{\bs p}
	&=	E_\tx{out}+\ol{\bs V}_\tx{out}\cdot\pn{\bs p-\bs P_\tx{out}},\\
\stin
	&=	{\sigma_1+\sigma_2\ov\pn{\bs V_1-\bs V_2}^2},&
\stout
	&=	{\sigma_3+\sigma_4\ov\pn{\bs V_3-\bs V_4}^2},
}
\al{
\sigma
	&=	{1\ov{1\ov\sigma_\tx{in}}+{1\ov\sigma_\tx{out}}}
	=	{1\ov{1\ov\sigma_1}+{1\ov\sigma_2}+{1\ov\sigma_3}+{1\ov\sigma_4}},&
\sigma_+
	&=	\sigma_\tx{in}+\sigma_\tx{out},\\
\vs
	&=	{1\ov{1\ov\vs_\tx{in}}+{1\ov\vs_\tx{out}}}
	=	{\stin\stout\ov\stin+\stout},&
\vs_+
	&=	\stin+\stout,
}
\al{
E_\sigma
	&=	\sqrt{M^2+\bs P_\sigma^2},&
\E_\sigma
	&=	\sqrt{E_\sigma^2-i\epsilon}
	=	\sqrt{M^2+\bs P_\sigma^2-i\epsilon},
}
\al{
\omega_\vs\fn{\bs p}
	&=	{\stin\omega_\tx{in}\fn{\bs p}+\stout\omega_\tx{out}\fn{\bs p}\ov\stin+\stout},&
E_\vs
	&=	{\stin E_\tx{in}+\stout E_\tx{out}\ov\stin+\stout},
}
\al{
\delta\ol{\bs\Xi}
	&=	\ol{\bs\Xi}_\tx{out}-\ol{\bs\Xi}_\tx{in},&
\delta\mf T
	&=	\mf T_\tx{out-int}-\mf T_\tx{in-int},
}
\al{
\Delta\T
	&=	\delta\mf T+\vs_+\Im\E_\sigma+i\stout\pn{E_\tx{out}-\Re\E_\sigma},
}
\al{
N	&=	\pn{-i\kappa}^2
		e^{-{\mc R_\tx{in}+\mc R_\tx{out}\ov2}}
			\prod_{a=1}^4{1\ov\sqrt{2E_a}}
			\pn{{1\ov\pi\sigma_a}}^{3/4},
}
\al{
\s_\tx{in}
	&:=	{1\ov{1\ov\stin}+{\ol{\bs V}_\tx{in}^2\ov\sigma_+}},\\
\s_\tx{out}
	&:=	{\stout\ov1+{\stout\ov\sigma_+}
			\pn{\ol{\bs V}_\tx{out}^2
				-{\s_\tx{in}\pn{\ol{\bs V}_\tx{in}\cdot\ol{\bs V}_\tx{out}}^2\ov\sigma_+}}},
}
\al{
\T_\tx{in-int}^\pm
	&:=	
		\s_\tx{in}\br{
			\sqbr{{\mf T_\tx{in-int}\ov\stin}
				+{\ol{\bs V}_\tx{in}\cdot\delta\ol{\bs\Xi}\ov\sigma_+}
				}
			-i\sqbr{\omega_\tx{in}\fn{0}
				\mp\E_\sigma
				+{\sigma\ol{\bs V}_\tx{in}\cdot\pn{\bs P_\tx{out}-\bs P_\tx{in}}\ov\sigma_\tx{in}}
					}
			},\\
\T_\tx{out-int}^\pm
	&:=	\s_\tx{out}\Bigg\{
			\sqbr{
				{\mf T_\tx{out-int}\ov\stout}
				+{\ol{\bs V}_\tx{out}\cdot\pn{
						\ol{\bs V}_\tx{in}\T_\tx{in-int}
						-\delta\ol{\bs\Xi}}
					\ov\sigma_+}
				}\nn
	&\phantom{:=\s_\tx{out}\Bigg\{}
			+i\sqbr{
				\omega_\tx{out}\fn{0}
				\mp\E_\sigma
					\pn{1-{\s_\tx{in}\ol{\bs V}_\tx{in}\cdot\ol{\bs V}_\tx{out}\ov\sigma_+}	
						}
				-{\sigma\ol{\bs V}_\tx{out}\cdot\pn{\bs P_\tx{out}-\bs P_\tx{in}}\ov\sigma_\tx{out}}
				}
			\Bigg\},
}
\al{
\ol{\bs V}_\sigma
	&:=	\sigma\pn{
			{\ol{\bs V}_\tx{in}\ov\sigma_\tx{in}}
			+{\ol{\bs V}_\tx{out}\ov\sigma_\tx{out}}
			},&
\ol{\bs V}_\vs
	&:=	
			{\stin\ol{\bs V}_\tx{in}+\stout\ol{\bs V}_\tx{out}\ov\stin+\stout},&
\mf T_\vs
	&:=	\vs\pn{{\mf T_\tx{in-int}\ov\stin}+{\mf T_\tx{out-int}\ov\stout}},
}
\al{
E_\vs
	&:=	{\vs_\tx{in}E_\tx{in}+\vs_\tx{out}E_\tx{out}\ov\vs_\tx{in}+\vs_\tx{out}},
}
\al{
\delta\omega\fn{\bs p}
	&:=	\omega_\tx{out}\fn{\bs p}-\omega_\tx{in}\fn{\bs p},\\
\delta\ol{\bs\Xi}
	&:=	\ol{\bs\Xi}_\tx{out}-\ol{\bs\Xi}_\tx{in},\\
\delta\mf T
	&:=	\mf T_\tx{out-int}-\mf T_\tx{in-int},\\
\DelT
	&:=	\delta\mf T+\stout\Im\E_\sigma+i\stout\pn{E_\tx{out}-\Re\E_\sigma}.
}

\section{Expressions for simplifying configuration}\label{simplifying section}
We list expressions for the simplifying configuration~\eqref{simple configuration}.
\begin{itemize}
\item Eqs.~\eqref{I star} and \eqref{V star}:
\al{
\mc I_\fs\fn{x^0,y^0}
	&\to
		\int{\df p^0\ov2\pi i}
			{e^{-ip^0\pn{y^0-x^0}}\ov
				-\pn{p^0}^2+\bs P_\sigma^2
				+M^2
				-{(\delta\ol{\bs\Xi})^2\ov\sigma_+^2}
				-i\pn{
					\epsilon
					-{2\bs P_\sigma\cdot\delta\ol{\bs\Xi}\ov\sigma_+}
					}
				},\\
V_\fs\fn{x^0,y^0}
	&\to
			-{\sigma\ov2}\pn{\bs P_\tx{out}-\bs P_\tx{in}}^2
			-{1\ov2\sigma_+}\pn{\ol{\bs\Xi}_\tx{out}-\ol{\bs\Xi}_\tx{in}}^2
			-i\ol{\bs\Xi}_\sigma\cdot\pn{\bs P_\tx{out}-\bs P_\tx{in}}
			-i\omega_\tx{in}\fn{0}x^0+i\omega_\tx{out}\fn{0}y^0.
}
\item
Eq.~\eqref{y0 left}:
\al{
\mc S
	&\to
		{\paren{2\pi\sigma}^{3/2}N\ov2\E_\sigma}
		e^{
			-{\sigma\ov2}\pn{\bs P_\tx{out}-\bs P_\tx{in}}^2
			-{1\ov2\sigma_+}\pn{\ol{\bs\Xi}_\tx{out}-\ol{\bs\Xi}_\tx{in}}^2
			-i\ol{\bs\Xi}_\sigma\cdot\pn{\bs P_\tx{out}-\bs P_\tx{in}}
			}\nn
	&\quad\times
		\int_{-\infty}^\infty\df y^0\,e^{
			-{1\ov2\stout}\paren{y^0-\mf T_\tx{out-int}}^2
			+iE_\tx{out}y^0
			-{\mf T_\tx{in-int}^2\ov2\stin}
			}\nn
	&\quad\times
		\sqrt{2\pi\stin}\left[
			e^{
			{1\ov2\stin}\pn{
				{\mf T_\tx{in-int}}
				+i\stin\Pn{\E_\sigma- E_\tx{in}}
				}^2
			-i\E_\sigma y^0
			}
			{1
			+\erf\fn{
				y^0
				-{\mf T_\tx{in-int}}
				-i\stin\Pn{\E_\sigma- E_\tx{in}}
					\ov\sqrt{2\stin}
				}
				\ov2}
				\right.\nn
	&\phantom{\quad\times\sqrt{2\pi\stin}\Bigg[}
				\left.\mbox{}
			+e^{
			{1\ov2\stin}\pn{
				{\mf T_\tx{in-int}}
				-i\stin\Pn{\E_\sigma+ E_\tx{in}}
				}^2
			+i\E_\sigma y^0
			}
			{1+\erf\fn{
				{\mf T_\tx{in-int}}
				-i\stin\Pn{\E_\sigma+ E_\tx{in}}
				-y^0
					\ov\sqrt{2\stin}
				}
				\ov2}
		\right].
}
\item
Eq.~\eqref{final full expression}:
\al{
S	&\to
		{\paren{2\pi\sigma}^{3/2}N\ov2\E_\sigma}
		e^{
			-{\sigma\ov2}\pn{\bs P_\tx{out}-\bs P_\tx{in}}^2
			-{1\ov2\sigma_+}\pn{\ol{\bs\Xi}_\tx{out}-\ol{\bs\Xi}_\tx{in}}^2
			-i\ol{\bs\Xi}_\sigma\cdot\pn{\bs P_\tx{out}-\bs P_\tx{in}}
			+iE_\tx{out}\mf T_\tx{out-int}
			-iE_\tx{in}\mf T_\tx{in-int}
			}
		\sqrt{2\pi\stin }\sqrt{2\pi\stout }\nn
	&\quad\times\Bigg[
		e^{
			-{\stin\pn{\E_\sigma-E_\tx{in}}^2\ov2}
			-{\stout\pn{E_\tx{out}-\E_\sigma}^2\ov2}
			-i\E_\sigma\delta\mf T
			}
		\pn{
			\theta\fn{\delta\mf T+\vs_+\Im\E_\sigma}
			+{\erf\fn{\Delta\T\ov\sqrt{2\stout}}-\sgn\fn{\Delta\T\ov\sqrt{2\stout}}\ov2}
			}\nn
	&\phantom{\quad\times\Bigg[}
		+e^{
			-{\stin\pn{\E_\sigma+ E_\tx{in}}^2\ov2}
			-{\stout\pn{E_\tx{out}+\E_\sigma}^2\ov2}
			+i\E_\sigma\delta\mf T
			}\nn
	&\phantom{\quad\times\Bigg[+}
	\times
			\pn{\theta\fn{-\delta\mf T+\vs_+\Im\E_\sigma}
			+{\erf\fn{-\delta\mf T+\vs_+\Im\E_\sigma-i\stout\pn{E_\tx{out}+\Re\E_\sigma}\ov\sqrt{2\stout}}-\sgn\fn{-\delta\mf T+\vs_+\Im\E_\sigma-i\stout\pn{E_\tx{out}+\Re\E_\sigma}\ov\sqrt{2\stout}}\ov2}
			}
		\Bigg].
		\label{final simplified}
}
\item
Eq.~\eqref{theta function first appeared}:
\al{
&\to\theta\fn{
		\delta\mf T
		+\vs_+\Im\E_\sigma
			}.
}
\item
Eq.~\eqref{theta functions}:
\al{
&\to\theta\fn{
		-\delta\mf T
		+\vs_+\Im\E_\sigma
			}.
}
\end{itemize}

\section{Details on Fig.~\ref{illustration}}\label{Details on Fig 2}
In Fig.~\ref{illustration}, we have plotted $\ab{\mc S}^2$
(up to the normalization factor~\eqref{factor C} below),
using Eq.~\eqref{emerging 1 to 2},
as a function of final state positions
\al{
x_3
	&:=	\ab{\bs X_3},&
x_4
	&:=	\ab{\bs X_4}.
}
We have focused on the back-to-back scattering in the center-of-mass frame,
\al{
\bs P_\tx{in}
	&:=	\bs P_1+\bs P_2=0,&
\bs P_\tx{out}
	&:=	\bs P_3+\bs P_4=0,&
\ab{\bs P_1}
	&=	\ab{\bs P_2}=\ab{\bs P_3}=\ab{\bs P_4}
	\quad\Pn{\,=\ab{\bs P}\,},
}
with the wave-packet positions being collinear to their momenta, $\bs X_a\propto\bs P_a$ ($=E_a\bs V_a$; $a=1,\dots,4$), and overlapping at the origin:
\al{
\bs\Xi_1
	&=	\bs X_1-\bs V_1\pn{-T}
	=	0,&
\bs\Xi_3
	&=	\pn{x_3-T}\bs V,\\
\bs\Xi_2
	&=	-\bs X_1-\pn{-\bs V_1}\pn{-T}
	=	0,&
\bs\Xi_4
	&=	\pn{-x_4+T}\bs V,
}
where
\al{
\bs V
	&:=	\bs V_3	
	=	-\bs V_4,&
\bs P
	&:=	\bs P_3
	=	-\bs P_4,&
\bigg(\,\bs V
	&=	{\bs P\ov\sqrt{m^2+\bs P^2}},\,\bigg)
}
and the initial and final times of the wave packets are taken to be $T_1=T_2=-T$ and $T_3=T_4=T$, respectively.
It follows that $\ab{\bs V_1}=\ab{\bs V_2}=\ab{\bs V_3}=\ab{\bs V_4}=\ab{\bs V}$.
The widths of the wave packets are chosen to be symmetric $\sigma_1=\sigma_2\ll\sigma_3=\sigma_4$.
As a result, we obtain
\al{
E_\tx{in}
	&=	2\sqrt{m^2+\bs P^2},&
E_\tx{out}
	&=	2\sqrt{m^2+\bs P^2},\\
\sigma_\tx{in}
	&=	{\sigma_1\ov2},&
\sigma_\tx{out}
	&=	{\sigma_3\ov2},\\
\ol{\bs V}_\tx{in}
	&=	0,&
\ol{\bs V}_\tx{out}
	&=	0,\\
\omega_\tx{in}\fn{\bs p}
	&=	E_\tx{in}
	=	2\sqrt{m^2+\bs P^2},&
\omega_\tx{out}\fn{\bs p}
	&=	E_\tx{out}
	=	2\sqrt{m^2+\bs P^2},\\
\stin
	&=	{\sigma_1\ov2\bs V^2},&
\stout
	&=	{\sigma_3\ov2\bs V^2},\\
\sigma
	&\simeq
		{\sigma_1\ov2},&
\sigma_+
	&\simeq
		{\sigma_3\ov2},\\
\vs
	&\simeq
		{\sigma_1\ov2\bs V^2},&
\vs_+
	&\simeq
		{\sigma_3\ov2\bs V^2},
}
\al{
\bs P_\sigma
	&=	0,&
E_\sigma
	&=	M,&
\E_\sigma
	&=	
		M-i{\epsilon\ov2M},&
\Im\E_\sigma
	&=
		-{\epsilon\ov2M},
}
\al{
\omega_\vs\fn{\bs p}
	&=	2\sqrt{m^2+\bs P^2},&
E_\vs
	&=	2\sqrt{m^2+\bs P^2},\\
\bs P_\tx{out}-\bs P_\tx{in}
	&=	0,&
E_\tx{out}-E_\tx{in}
	&=	0,\\
\delta\ol{\bs\Xi}
	&=	\ol{\bs\Xi}_\tx{out}
	=	{x_3-x_4\ov2}\bs V,&
\delta\mf T
	&=	\mf T_\tx{out-int}
	=	T-{x_3+x_4\ov2},
}
\al{
\Delta\T
	&\simeq
		T-{x_3+x_4\ov2}
		-{\epsilon\sigma_3\ov4M\bs V^2}
		+i{\sigma_3\ov2\bs V^2}\pn{2\sqrt{m^2+\bs P^2}-M},
		\label{Delta T evaluated}
}
\al{
N	&=	-\kappa^2
			{1\ov\pn{2E_1}^2}
			\pn{{1\ov\pi\sigma_1}}^{3/2}
			\pn{{1\ov\pi\sigma_3}}^{3/2},
}
where we have taken a small $\epsilon$ and
\al{
\ol{\bs\Xi}_\tx{in}
	&=	0,&
\ol{\bs\Xi}_\tx{out}
	&=	{x_3-x_4\ov2}\bs V,\\
\mf T_\tx{in-int}
	&=	0,&
\mf T_\tx{out-int}
	&=	T-{x_3+x_4\ov2},\\
\mc R_\tx{in}
	&=	0,&
\mc R_\tx{out}
	&=	0.
}
The resultant expression for $\ab{\mc S}^2$ is
\al{
\ab{\mc S}^2
	&=	C
		{M^4\ov\ab{\bs P}^4}
		e^{
			-{\bs V^2\ov2\sigma_3}\pn{x_3-x_4}^2
			-{\sigma_3\ov2\bs V^2}\pn{2\sqrt{m^2+\bs P^2}-M}^2
			}\nn
	&\quad\times
		\ab{
			\theta\fn{T-{x_3+x_4\ov2}-{\epsilon\sigma_3\ov4M\bs V^2}}
			+{\erf\fn{{\ab{\bs V}\DelT\ov\sqrt{\sigma_3}}}
				-\sgn\fn{{\ab{\bs V}\DelT\ov\sqrt{\sigma_3}}}\ov2}
			}^2,
			\label{plotted S sqaured}
}
with $\DelT$ being given by Eq.~\eqref{Delta T evaluated} and
\al{
C	&:=	{\kappa^4\ov64\pi M^6}{\sigma_1\ov\sigma_3^2}
		\label{factor C}
}
is a constant factor.
In FIG.~2 in the main text, we have plotted $\ab{\mc S}^2/C$.

The plot is for the parameters
\al{
m	&=	{M\ov4},&
T	&=	{100\ov M},&
\epsilon
	&=	0.01M^2,&
\sigma_3
	&=	{10\ov M^2}.
}
On the upper panels, we have chosen the momenta to be on the resonance:
\al{
2\sqrt{m^2+\bs P^2}=M,
}
namely, $\ab{\bs P}={\sqrt{3}\ov4}M$.
On the lower panels, we have chosen them to be off the resonance: $\ab{\bs P}={M\ov4}$.

On the left panels, we include only the first term in the absolute squared in Eq.~\eqref{plotted S sqaured}, namely,
\al{
\propto	\br{\theta\fn{T-{x_3+x_4\ov2}-{\epsilon\sigma_3\ov4M\bs V^2}}}^2.
}
On the right panels, we take into account all the contributions including the interference with the boundary contribution
\al{
\propto
	\theta\fn{T-{x_3+x_4\ov2}-{\epsilon\sigma_3\ov4M\bs V^2}}
	\pn{{\erf\fn{{\ab{\bs V}\DelT\ov\sqrt{\sigma_3}}}
				-\sgn\fn{{\ab{\bs V}\DelT\ov\sqrt{\sigma_3}}}\ov2}
+\tx{h.c.}}
	\label{interference}
}
and the boundary-only contribution
\al{
\propto\ab{\erf\fn{{\ab{\bs V}\DelT\ov\sqrt{\sigma_3}}}
				-\sgn\fn{{\ab{\bs V}\DelT\ov\sqrt{\sigma_3}}}\ov2}^2.
				\label{boundary only}
}
The interference~\eqref{interference} contributes at $\delta\mf T=T-{x_3+x_4\ov2}>0$ ($x_3+x_4<2T$) negatively on the resonance (upper-right panel) and mainly positively off the resonance (lower-right panel). The boundary-only contribution~\eqref{boundary only} exists at both signs of $\delta\mf T$.

\bibliographystyle{JHEP}
\bibliography{refs}

\end{document}